# Spectrum Sharing in STAR-RIS-assisted UAV with NOMA for Cognitive Radio Networks

Ali Nazari, Ali Olfat, *Member, IEEE*
Signal Processing and Communication System Laboratory, University of Tehran, Tehran, Iran.
e-mails: ali.nazary@ut.ac.ir and aolfat@ut.ac.ir

***Abstract*—As an emerging technology, the simultaneous transmitting and reflecting reconfigurable intelligent surface (STAR-RIS) can improve the spectrum efficiency (SE) of primary users (PUs) and secondary users (SUs) in cognitive radio (CR) networks by mitigating the interference of the incident signals. The STAR-RIS-assisted unmanned aerial vehicle (UAV) can fully cover the dynamic environment through high mobility and fast deployment. According to the dynamic air-to-ground channels, the STAR-RIS-assisted UAV may face a challenge configuring their elements' coefficients (i.e., reflecting and transmitting the amplitude and phases). Hence, to meet the requirements of dynamic channel determination with the SE approach, this paper proposes the sum rate maximization of both PUs and SUs through non-orthogonal multiple access in CR network to jointly optimize the trajectory and transmission-reflection beamforming design of the STAR-RIS-assisted UAV, and power allocation. Since the non-convex joint optimization problem includes coupled optimization variables, we develop an alternative optimization algorithm. Simulation results study the impact of: 1) the significant parameters, 2) the performance of different intelligence surface modes and STAR-RIS operating protocols, 3) the joint trajectory and beamforming design with fixed and mobile users, and 4) STAR-RIS capabilities such as mitigating the interference, and how variations in the roles of elements dynamically.**

## I. Introduction

Reconfigurable intelligent surfaces (RIS) are a promising technology with high data rates, lower latency, and massive connectivity abilities to enhance the network performance of fifth generation (5G) and beyond. The low cost, passive controllable RIS's elements improve signal quality and coverage by reconfiguring radio environments [1]. Due to the potential of controlling the amplitude and phase of the incident signals, RIS plays an important role in optimizing spectrum sharing in cognitive radio (CR) networks by alleviating the interference signals at secondary users [2]. Nevertheless, since the RIS only reflects the incident signals, it attracts attention in scenarios where users are placed on the same side as the RIS. Therefore, to enhance the spectral efficiency in optimizing the wireless link between transceivers, we employ the new generation of intelligence surfaces that simultaneously reflect and transmit incident signals [3]. The simultaneously transmitting and reflecting RIS (STAR-RIS) technology which is also referred to as intelligent omni-surface (IOS), separates the incident signal into reflecting and transmitting signals to intelligently control their coefficients simultaneously [4]. Hence, the 360° coverage is the achievement of this technology in dynamic environments. The RIS-assisted CR communication network including maximization of achievable rate within optimizing the secondary power transmission and RIS reflection beamforming is investigated in [5]. The authors considered the spectrum sharing aspect of RIS-assisted vehicular network to jointly maximize sum capacity of vehicle-to-infrastructure by ensuring the reliability of vehicle-to-vehicle communications in [6].

According to the growing demand for spectrum sharing and the probability of path blockage, as well as the presence of weak non-line-of-sight (NLOS) and line-of-sight (LOS) links, investigating the deployment of STAR-RIS is a promising and challenging study within 3rd Generation Partnership Project (3GPP) Release-19 standards aimed at enhancing network throughput. To overcome this challenge, we assume that STAR-RIS is mounted on the unmanned aerial vehicle (UAV) to illustrate the mobile STAR-RIS effectiveness on the CR network performance from the perspective of spectrum sharing among primary users (PUs) and secondary users (SUs). Due to the high mobility and flexibility of UAVs to cover remote or disaster areas, UAV-assisted STAR-RIS can fully cover the users on the same side and opposite side of the surface. It is worth noting that UAVs as relays or base stations (BS) in STAR-RIS-assisted scenarios are one of the authors' interests in improving the network's performance. In [7], the authors focus on maximizing the average rate of a ground node using a UAV and the reflection/transmission elements of an IOS through a joint optimization of the UAV's trajectory and the IOS's passive beamforming. The authors in [8], assumed STAR-RIS assisted UAV's multiple antennae to maximize total rate within joint optimization of the UAV's beamforming trajectory design and the STAR-RIS's passive beamforming. The maximization of the system throughput by optimizing the beamforming vectors of STAR-RIS in the role of a relay, the trajectory of UAV as an aerial BS, and power allocation is studied in [9]. The [10] proposed max-min achievable rate within optimal approach to find the scheduling of users, phase shifts of IOS, and trajectory design of UAV. However, our main goal is to investigate the impact of mounting STAR-RIS on UAVs as a dynamic scenario. The authors in [11], investigate UAV path planning to maximize data rate by amplifying the transmitting and reflecting signals due to a specific user's location region, i.e., same and opposite sides of STAR-RIS, respectively. In [12], STAR-RIS mounted on a UAV to optimize the beamforming vectors and transmitted and reflected factors. However, in [12], the UAV flies over a predefined location, and adjusting the reconfigurable elements role toward the users is fixed during the flying time. Regarding the mobility of STAR-RIS mounted on the UAV toward the user's position, the elements' role of the intelligence surface changes during communication time. To the best of our knowledge, the related scenarios involving STAR-RIS-assisted UAVs haven't been studied this concern yet. Therefore, not only practical scenarios shouldn't require the predefined location of users, and the element's role, but also should be dynamically assigned elements during flight, depending on the user's real-time location relative to the STAR-RIS (i.e., whether the user is located on the same or opposite sides).

In CR Networks, SUs are served under the quality of service (QoS) constraint of PUs. The SUs ensure that their spectrum access does not interfere or degrade the QoS of PUs. In order to improve spectrum sharing in CR network with SUs and PUs, we adopt power-domain non-orthogonal multiple access (PD-NOMA) as a transmission technique. PD-NOMA significantly improves spectrum efficiency by employing the superposition of power-domain resources in a non-orthogonal manner [13]. In addition, successive interference cancellation (SIC) is exploited at the receivers to alleviate the interference. Integrating STAR-RIS with the PD-NOMA technique in CR network enhances system performance by adjusting the channel gains of PUs and SUs to improve spectral efficiency. As well as STAR-RIS capability can develop the received interference by modifying the SIC. In [14], the authors examined the balance between spectrum efficiency (SE) and energy efficiency (EE) in RIS-assisted CR network using NOMA. In [15] is studied STAR-RIS aided downlink NOMA multi-cell network to derive the coverage probability and ergodic rate of user equipments. STAR-RIS assisted MISO-NOMA with queue awareness is considered in [16], which maximizes the per-slot queue-weighted sum rate by optimizing active and passive beamforming within each time slot. An IOS-assisted NOMA downlink system incorporating spatially correlation in channels and phase error, was proposed in [17]. A wireless powered NOMA system leveraging STAR-RIS technology and demonstrating the outage probability and sum of throughput performance is studied in [18]. In STAR-RIS assisted NOMA system, the joint beamforming and position design of weighted sum-rate maximization problem subject to the particular SIC decoding order, transmission/reflection power constraints, and user's QoS is investigated in [19]. In [20], the authors considered a STAR-RIS assisted NOMA system maximization of the achievable total rate through the joint optimization of SIC order, power allocation, BS beamforming, and STAR-RIS beamforming.

Consequently, motivated by the aforementioned contribution of STAR-RIS-assisted UAV in CR network, the performance of both PUs and SUs is enhanced through jointly designing the channel gain and spectrum sharing. This is achieved by simultaneously transmitting and reflecting incident signals. The authors recently proposed a similar scenario where a UAV-carried RIS for CR network in [21]. They designed the UAV trajectory for average rate maximization of SU under the quality of PU's service.

In this study attempts to express the challenges of STAR-RIS-assisted UAV and subsequently address them. We design a new framework for a STAR-RIS-assisted UAV in CR network with PD-NOMA technique. According to spectral deficiency, the STAR-RIS-assisted UAV significantly improves spectrum sharing by information transmission and interference management. We consider a pair of device-to-device Internet of Things (D2D-IoT) devices as SUs who communicate with each other through a STAR-RIS-assisted UAV under the QoS constraint of PUs. By achieving the underlay mode as a spectrum-sharing paradigm, the SU pair simultaneously shares the same band with PUs and controls their mutual interference [22]. In order to fully coverage and improve the performance of both PUs and SUs from a spectrum-sharing perspective, the optimization problem jointly investigates the trajectory of the STAR-RIS-assisted UAV, transmission-reflection beamforming design, and power allocation. Our additional contributions to the proposed system model are summarized as follows:

- We consider a general and practical system model where the set of PUs with weak or blocked connections to the BS is served by STAR-RIS-assisted UAV using the PD-NOAM technique, and a pair of SUs communicates through STAR-RIS-assisted UAV simultaneously, too.
- According to STAR-RIS-assisted UAV generates dynamic environments, we evaluate the impact of mobility on STAR-RIS's elements' role specifically as a transmitter or reflector during communication time. In this novel scenario of a STAR-RIS-assisted UAV CR network, we continuously specify the optimal elements' role for the first time, and then reconfigure their amplitude and phases during flying time.
- In addition, PD-NOMA is regarded as a SE improvement technique. Due to the dynamic SIC ordering, the STAR-RIS-assisted UAV should also monitor the environment to adjust the incident signals, eliminate interference, and enhance the overall sum rate.
- Aiming at sum rate maximization of both PUs and SUs the joint optimization problem incorporates UAV's trajectory constraints, the QoS constraint for PUs under SUs' spectrum sharing utilization, STAR-RIS's design strategy of transmission and reflection elements, and power transmission.
- Based on the proposed non-convex and complicated joint optimization problem and coupled optimization variables, we develop an alternative optimiza-tion (AO) algorithm. This approach employs Geometric Programming (GP) for the trajectory design of the STAR-RIS-assisted UAV, then applies semi-definite relaxation (SDR) for the transmission-reflection beamforming design of the STAR-RIS-assisted UAV, and finally utilizes the Difference-of-Two-Concave Functions (D.C.) approximation for the power allocation.

- The simulation results study the impact of effective parameters such as the number of elements, the number of PUs, and SUs' pairs, the communication time of trajectory design, and the power budget on the performance of the proposed system model. Investigating different modes of intelligence surface-assisted UAV and the STAR-RIS operating protocols is demonstrated, too. To study the PD-NOMA transmission technique's effectiveness in spectrum-sharing improvement, the orthogonal multiple access (OMA) technique is employed as a comparison method. In addition, motivated by the proposed contribution we present the requirement of the dynamic energy splitting ratio of optimization elements with fixed and mobile PUs and SUs.

*Notations:* In this paper, scalars are defined with italic letters. Vectors and matrices are denoted as boldface lowercase and uppercase letters, respectively. The notation $\mathbb{R}^{M\times 1}$ indicates a vector space consisting of $M$-dimensional real-valued, while $\mathbb{C}^{M\times N}$ refers to the space of $M \times N$-dimensional complex-valued. The transpose, and conjugate transpose of vector $\mathbf{x}$, are represented as $\mathbf{x}^T$, and $\mathbf{x}^H$, respectively. The magnitude of a complex number is indicated by $|.|$ and the Euclidean norm of a vector is $\|\ .\ \|$, too. $\mathbb{E}\{x\}$, and $\mathrm{diag}(\mathbf{x})$ represent the expectation of $x$, and the diagonal matrix, respectively. The complex Gaussian distribution with mean of $\mu$ and variance of $\sigma^2$ is denoted by $\mathcal{CN}(\mu, \sigma^2)$. Furthermore, $\mathbf{X} \succeq 0$ indicates that matrix $\mathbf{X}$ is positive semi-definite

## II. System Model

### A. Scenario Description

We consider a STAR-RIS employing the PD-NOMA technique in CR network, where the direct links between BS and PUs are too weak and insignificant as illustrated in Fig 1. Therefore, we adopt STAR-RIS-assisted UAV, to assist the communication links, and improve the spectrum efficiency. Additionally, there is no direct link between the D2D-IoT devices as SU pair $\{s_1, s_2\}$, and they attempt to communicate with each other through the STAR-RIS-assisted UAV. All transceivers such as BS, PUs, and SUs have a single antenna. The set of PUs is defined as $k \in \mathcal{K} = \{1, 2, \ldots, K\}$, where $|\mathcal{K}| = K$ denoted the total number of PUs, and $p_k$ represents the $k$-th PU. The location of BS is given by $\mathbf{w}_B = [x_B, y_B]^T \in \mathbb{R}^{2\times 1}$ and the location of both PUs and SUs represented by $\mathbf{r}_u = [x_u, y_u]^T \in \mathbb{R}^{2\times 1}$, where $u \in \mathcal{U} = \{p_k, s_1, s_2 \mid k \in \mathcal{K}\}$.

Assuming the STAR-RIS-assisted UAV is flying above the ground at a fixed altitude $H$ [23], its horizontal

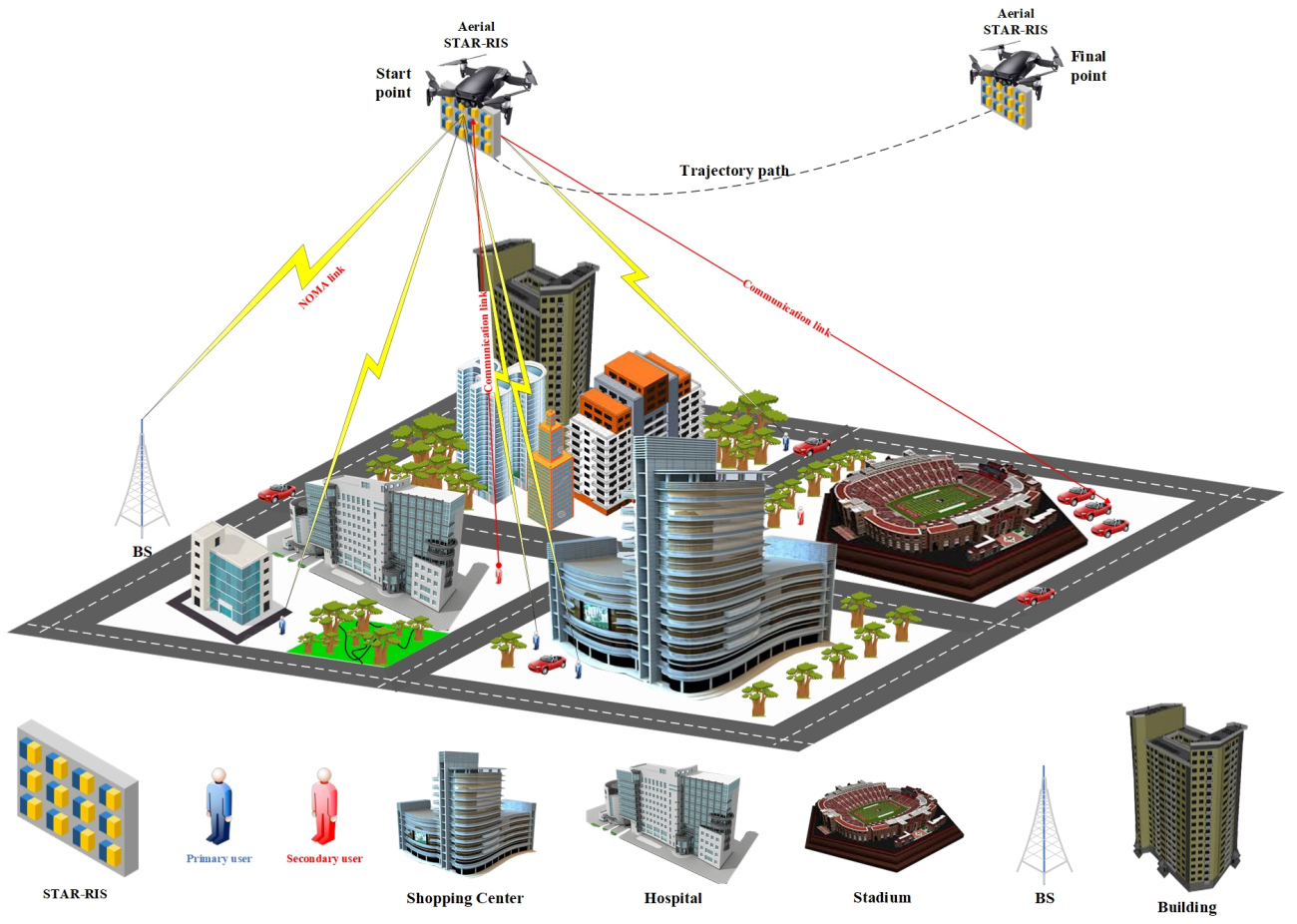

Fig. 1: The considered system model.

Cartesian coordinates at time $t$ is represented as $\mathbf{q}(t) = [x(t), y(t)]^T \in \mathbb{R}^{2\times 1}$. Here, STAR-RIS-assisted UAV starts the mission from $\mathbf{q}(0)$ toward the final point $\mathbf{q}(T)$ at the end of the communication period $T$. To ease exposition of the trajectory design of STAR-RIS-assisted UAV, the communication time $T$ is discretized into $N$ equal time slots, indexed by $n \in \mathcal{N} = \{0, 1, \ldots, N\}$. For $n = 0$, and $n = N$ the start and final points are $\mathbf{q}[0]$, $\mathbf{q}[N]$, respectively. Therefore, the horizontal location of STAR-RIS-assisted UAV in each time slot $n$ is $\mathbf{q}[n] = [x_q[n], y_q[n]]^T$. Furthermore, we define $\mathbf{Q} = \{\mathbf{q}[n] | n \in \mathcal{N}\}$ for a STAR-RIS-assisted UAV trajectory design which restricts between two sequential time slots as the following:

$$\|\mathbf{q}[n] - \mathbf{q}[n-1]\| \leqslant V_{\max}\delta, \ n = 1, \ldots, N, \tag{1}$$

where $V_{\max}\delta$ is the maximum horizontal distance that the UAV can travel at its maximum speed $V_{\max}$ in each time slot $n$, and the length of each elemental time slot $\delta = \frac{T}{N}$ is sufficiently small to ensure the UAV's position is almost steady throughout each time slot $n$.

### B. STAR-RIS Structure

This section details the structure of STAR-RIS which comprises $M$ reconfigurable elements. Based on the literature [24]–[26], three practical protocols are available: energy splitting (ES), mode switching (MS), and time switching (TS). This study focuses on the ES protocol, where the incident signal's energy is split into transmitting and reflecting signals with distinct energy levels. Consequently, all the STAR-RIS elements i.e. $M$ are assumed to operate simultaneously in both transmission and reflection modes by assigning an energy-splitting ratio. The wireless signal incident $s_m$ is transmitted from BS to the $m$-th element of STAR-RIS, then reconfigured by the corresponding transmission and reflection coefficients by the $m$-th element, where $m \in \mathcal{M} = \{1, \ldots, M\}$, and $|\mathcal{M}| = M$ denoted the total number of elements. Respect to the incident signal $s_m$, $s_m^r$ is the reflecting signal in the same space, although $s_m^t$ is the transmitting signal in the opposite space, and expressed respectively as follows:

$$s_m^r = \left(\sqrt{\alpha_m^r} e^{j\theta_m^r}\right) s_m, \tag{2a}$$

$$s_m^t = \left(\sqrt{\alpha_m^t} e^{j\theta_m^t}\right) s_m, \tag{2b}$$

where $\alpha_m^t, \alpha_m^r \in [0,1]$ are the amplitude adjustment coefficients, and $\theta_m^t, \theta_m^r \in [0, 2\pi)$ are phase shift coefficients of the $m$-th element for transmission and reflection. According to the law of energy conservation [26], all the elements must be satisfied the following constraint:

$$\alpha_m^t + \alpha_m^r = 1, \;\forall m \in \mathcal{M}. \tag{3}$$

The beamforming vector $\boldsymbol{\varphi}^{\varkappa}[n]$ at time slot $n$ is defined as $\boldsymbol{\varphi}^{\varkappa}[n] = [\sqrt{\alpha_1^{\varkappa}[n]}e^{j\theta_1^{\varkappa}[n]}, \sqrt{\alpha_2^{\varkappa}[n]}e^{j\theta_2^{\varkappa}[n]}, \ldots, \sqrt{\alpha_M^{\varkappa}[n]}e^{j\theta_M^{\varkappa}[n]}]^H \in \mathbb{C}^{M\times 1}$, where $\varkappa \in \Xi = \{t, r\}$ indicates the element's mode, corresponding to either transmission or reflection. Additionally, $\boldsymbol{\Theta}^{\varkappa}[n] = \mathrm{diag}\left(\boldsymbol{\varphi}^{\varkappa}[n]\right) \in \mathbb{C}^{M\times M}$ is the corresponding diagonal beamforming matrix, and $\boldsymbol{\Phi} = \{\boldsymbol{\varphi}^{\varkappa}[n] | \; \varkappa \in \Xi, \; n = 1, \ldots, N\}$ shows the set of transmission and reflection beamforming vector for all time slot.

### C. Channel Gain

It is assumed that the communication links from the BS to the STAR-RIS-assisted UAV, and from the STAR-RIS-assisted UAV to the PUs and SUs, follow the free-space path-loss model [27]. Therefore, the channel gain from BS to STAR-RIS-assisted UAV at time slot $n$ with $\beta_0$ as the channel gain at a reference distance $d_0 = 1$ m, is expressed as:

$$\mathbf{g}[n] = \sqrt{\beta_0 d_{BI}^{-\alpha}[n]}\,\bar{\mathbf{g}}[n], \tag{4}$$

where $\bar{\mathbf{g}}[n] = [1, e^{-j\frac{2\pi}{\lambda}d\omega_{BI}[n]}, \ldots, e^{-j\frac{2\pi}{\lambda}d(M-1)\omega_{BI}[n]}]^T \in \mathbb{C}^{M\times 1}$, $\omega_{BI}[n] = \frac{x_q[n]-x_B}{d_{BI}[n]}$ shows the cosine of the angle of arrival (AoA) of the signal from BS to the STAR-RIS-assisted UAV in time slot $n$, $d$ is the antenna separation, and $\lambda$ is the carrier wavelength. Also, $d_{BI}[n] = \sqrt{\|\mathbf{q}[n] - \mathbf{w}_B\|^2 + H^2}$ is the distance between BS and STAR-RIS-assisted UAV at time slot $n$.

Similarly, the channel gain from STAR-RIS-assisted UAV to PUs and SUs at time slot $n$ is given by:

$$\mathbf{v}_u[n] = \sqrt{\beta_0 d_{Iu}^{-\alpha}[n]}\,\bar{\mathbf{v}}_u[n], \tag{5}$$

where $\bar{\mathbf{v}}_u[n] = [1, e^{-j\frac{2\pi}{\lambda}d\omega_{Iu}[n]}, \ldots, e^{-j\frac{2\pi}{\lambda}d(M-1)\omega_{Iu}[n]}]^T \in \mathbb{C}^{M\times 1}$, $\omega_{Iu}[n] = \frac{x_q[n]-x_u}{d_{Iu}[n]}$ represents the cosine of the angle of departure (AoD) of the signal from STAR-RIS-assisted UAV to user $u$ in time slot $n$, and $d_{Iu}[n] = \sqrt{\|\mathbf{q}[n] - \mathbf{r}_u\|^2 + H^2}$ is the distance between STAR-RIS-assisted UAV and user $u$ at time slot $n$.

### D. Transmission Technique and Achievable Rate

Based on the PD-NOMA concept as an efficient transition technique, BS transmits the superpositioned signals $x_p[n]$ through STAR-RIS-assisted UAV to PUs at time slot $n$:

$$x_p[n] = \sum_{k=1}^{K} \sqrt{p_{p_k}[n]}\,x_{p_k}[n], \tag{6}$$

where $x_{p_k}[n]$ is requested signal by $k$-th PU at time slot $n$ with power transmission $p_{p_k}[n]$. In addition, the first secondary IoT device, $s_1$, transmits the signal $x_s[n]$ to the second secondary IoT device, $s_2$, through a STAR-RIS-assisted UAV at time slot $n$ with power transmission $p_{s_2}[n]$:

$$x_s[n] = \sqrt{p_{s_2}[n]}\,x_{s_2}[n], \tag{7}$$

where $x_{s_2}[n]$ is the signal requested by $s_2$ from $s_1$. We assume $\mathbf{P} = \{p_{p_k}[n] | \forall k \in \mathcal{K}, \; n = 1, \ldots, N\}$, and $E\{|x_{p_k}[n]|^2\} = E\{|x_{s_2}[n]|^2\} = 1, \;\forall k \in \mathcal{K}$.

The received signal at PUs and $s_2$ in time slot $n$ are expressed as follows, respectively:

$$\begin{aligned} y_{p_k}[n] = &\left(\mathbf{v}_{p_k}^H[n]\boldsymbol{\Theta}^{\varkappa}[n]\mathbf{g}[n]\right) x_p[n] \\ &+ \left(\mathbf{v}_{p_k}^H[n]\boldsymbol{\Theta}^{\varkappa}[n]\mathbf{v}_{s_1}[n]\right) x_s[n] + n_{p_k}, \end{aligned} \tag{8}$$

$$\begin{aligned} y_{s_2}[n] = &\left(\mathbf{v}_{s_2}^H[n]\boldsymbol{\Theta}^{\varkappa}[n]\mathbf{v}_{s_1}[n]\right) x_s[n] \\ &+ \left(\mathbf{v}_{s_2}^H[n]\boldsymbol{\Theta}^{\varkappa}[n]\mathbf{g}[n]\right) x_p[n] + n_{s_2}, \end{aligned} \tag{9}$$

where $n_{p_k} \sim \mathcal{CN}\left(0, \sigma^2\right)$, and $n_{s_2} \sim \mathcal{CN}\left(0, \sigma^2\right)$ are Additive White Gaussian Noise (AWGN) with variance $\sigma^2$.

Therefore, the achievable rate for each PU in time slot $n$ is shown as $R_{p_k}[n]$ in formulation (10). By employing SIC in PD-NOMA-based communication, the set $\xi[n] = \{|h_{BIp_w}[n]|^2 > |h_{BIp_k}[n]|^2 | \;\forall\; w, \; k \in \mathcal{K}\}$ represents the channel power gain ordering. This indicates that PU with a strong channel power gain can successfully remove the interference caused by the other PUs with weaker channel power gain, where $h_{BIp_k}[n] = \mathbf{v}_{p_k}^H[n]\boldsymbol{\Theta}^{\varkappa}[n]\mathbf{g}[n]$.

Additionally, the achievable rate of $s_2$ in each time slot $n$, is given by:

$$R_s[n] = \log_2\left(1 + \frac{p_{s_2}[n]\left|\mathbf{v}_{s_2}^H[n]\boldsymbol{\Theta}^{\varkappa}[n]\mathbf{v}_{s_1}[n]\right|^2}{\sum_{k=1}^{K} p_{p_k}[n]\left|\mathbf{v}_{s_2}^H[n]\boldsymbol{\Theta}^{\varkappa}[n]\mathbf{g}[n]\right|^2 + \sigma^2}\right). \tag{11}$$

## III. Problem Formulation

To enhance spectrum sharing efficiency in STAR-RIS-assisted UAV within CR network using the PD-NOMA technique, we propose a joint problem formulation that optimizes the trajectory $\mathbf{Q}$, transmission-reflection beamforming design of the STAR-RIS-assisted UAV $\boldsymbol{\Phi}$, and power allocation $\mathbf{P}$, as follows:

$$\max_{\mathbf{Q},\boldsymbol{\Phi},\mathbf{P}} \sum_{n=1}^{N}\left(\sum_{k=1}^{K} R_{p_k}[n] + R_s[n]\right) \tag{12a}$$

$$\text{s.t.} \;\; \|\mathbf{q}[n] - \mathbf{q}[n-1]\| \leqslant V_{\max}\delta, \;\; n = 1, \ldots, N, \tag{12b}$$

$$R_{p_k}[n] \geqslant R^{\text{rsv}}, \;\forall k \in \mathcal{K}, \;\; n = 1, \ldots, N, \tag{12c}$$

$$\alpha_m^{\varkappa}[n] \in [0, 1], \;\forall \varkappa \in \Xi, \;\forall m \in \mathcal{M}, \;\; n = 1, \ldots, N, \tag{12d}$$

$$\theta_m^{\varkappa}[n] \in [0, 2\pi), \;\forall \varkappa \in \Xi, \;\forall m \in \mathcal{M}, \;\; n = 1, \ldots, N, \tag{12e}$$

$$\alpha_m^t[n] + \alpha_m^r[n] = 1, \;\forall m \in \mathcal{M}, \;\; n = 1, \ldots, N, \tag{12f}$$

$$\sum_{k=1}^{K} p_{p_k}[n] \leqslant P_{\max}, \;\; n = 1, \ldots, N, \tag{12g}$$

where based on constraint (12b), the maximum distance limitation between two sequential points is guaranteed,

$$R_{p_k}[n] = \log_2\left(1 + \frac{p_{p_k}[n]\left|\mathbf{v}^H_{p_k}[n]\mathbf{\Theta}^{\varkappa}[n]\mathbf{g}[n]\right|^2}{\sum_{w\in\xi[n]} p_{p_w}[n]\left|\mathbf{v}^H_{p_k}[n]\mathbf{\Theta}^{\varkappa}[n]\mathbf{g}[n]\right|^2 + p_{s_2}[n]\left|\mathbf{v}^H_{p_k}[n]\mathbf{\Theta}^{\varkappa}[n]\mathbf{v}_{s_1}[n]\right|^2 + \sigma^2}\right), \tag{10}$$

(12c) represents the QoS constraint with $R^{\rm rsv}$ as the minimum rate requirement, constraints (12d) and (12e) model the amplitude and phase limitation of STAR-RIS for both transmitter and reflector mode, and (12f) refers to amplitude constraint for the assumption of ES protocol. In order to show the maximum power budget $P_{\max}$ for serving all PUs, constraint (12g) is denoted.

The proposed problem (12) with coupled optimization variables $\mathbf{Q}$, $\mathbf{\Phi}$, and $\mathbf{P}$, and the presence of a non-convex objective function (12a), and non-convex constraint (12c) results in challenges to achieve an optimal solution.

## IV. Solution

In this section, the proposed non-convex optimization problem (12) becomes more tractable by adopting the AO algorithm. Accordingly, the proposed optimization problem (12) is decomposed into three sub-problems. Initially, the trajectory design of the STAR-RIS-assisted UAV, represented as $\mathbf{Q}$, is optimized using GP technique with given the transmission-reflection beamforming factor $\mathbf{\Phi}$ and power allocation $\mathbf{P}$. Next, the derived trajectory design of the STAR-RIS-assisted UAV and the given power allocation $\mathbf{P}$ are utilized to optimize the beamforming design, using SDR technique. Finally, the optimization variables obtained from the previous two steps, i.e., $\mathbf{Q}$, and $\mathbf{\Phi}$ are are exploited optimizing the power-allocation $\mathbf{P}$ by D.C. approximation. The optimization of three sub-problem is represented as:

$$\underbrace{\mathbf{Q}^0 \rightarrow \mathbf{\Phi}^0 \rightarrow \mathbf{P}^0}_{\text{Initialization}}, \ldots, \underbrace{\mathbf{Q}^t \rightarrow \mathbf{\Phi}^t \rightarrow \mathbf{P}^t}_{\text{Iteration } t}, \ldots, \underbrace{\mathbf{Q}^* \rightarrow \mathbf{\Phi}^* \rightarrow \mathbf{P}^*}_{\text{Optimal solution}}.$$

### *A. STAR-RIS-assisted UAV Trajectory Design Optimization*

The sub-problem of the STAR-RIS-assisted UAV trajectory design for a given transmission-reflection beamforming factor $\mathbf{\Phi}$, along with the corresponding power allocation $\mathbf{P}$, is proposed as:

$$\max_{\mathbf{Q}} \sum_{n=1}^{N-1}\left(\sum_{k=1}^{K} R_{p_k}[n] + R_s[n]\right) \tag{13a}$$

$$\text{s.t.} \quad \|\mathbf{q}[n] - \mathbf{q}[n-1]\| \leqslant V_{\max}\delta, \;\; n = 1, \ldots, N, \tag{13b}$$

$$R_{p_k}[n] \geqslant R^{\rm rsv}, \qquad \forall k \in \mathcal{K}, \;\; n = 1, \ldots, N-1, \tag{13c}$$

regarded to the objective function (13a), and the constraint (13c), the sub-problem (13) is non-convex. On the other hand, the defined channel power gain results in a complicated rate function. Therefore, we present the upper bound for the channel power gain to make a more tractable convexity of (13). The channel gain from BS to $p_k$, is rewritten as:

$$\mathbf{v}^H_{p_k}[n]\mathbf{\Theta}^{\varkappa}[n]\mathbf{g}[n] = \frac{\beta_0 \sum_{m=1}^{M} \sqrt{\alpha^{\varkappa}_m[n]}\, e^{j(\frac{2\pi d(m-1)}{\lambda}(\omega_{Ip_k}[n]-\omega_{BI}[n])-\theta^{\varkappa}_m[n])}}{\sqrt{d^{\alpha}_{Ip_k}[n] d^{\alpha}_{BI}[n]}}. \tag{14}$$

In addition the norm function of $\mathbf{v}^H_{p_k}[n]\mathbf{\Theta}^{\varkappa}[n]\mathbf{g}[n]$ is:

$$\begin{aligned}
|\mathbf{v}^H_{p_k}[n]\mathbf{\Theta}^{\varkappa}[n]\mathbf{g}[n]| &= \frac{\left|\beta_0 \sum_{m=1}^{M} \sqrt{\alpha^{\varkappa}_m[n]}\, e^{j(\frac{2\pi d(m-1)}{\lambda}(\omega_{Ip_k}[n]-\omega_{BI}[n])-\theta^{\varkappa}_m[n])}\right|}{\sqrt{d^{\alpha}_{Ip_k}[n] d^{\alpha}_{BI}[n]}} \\
&\leqslant \frac{|\beta_0| \sum_{m=1}^{M} \left|\sqrt{\alpha^{\varkappa}_m[n]}\, e^{j(\frac{2\pi d(m-1)}{\lambda}(\omega_{Ip_k}[n]-\omega_{BI}[n])-\theta^{\varkappa}_m[n])}\right|}{\sqrt{d^{\alpha}_{Ip_k}[n] d^{\alpha}_{BI}[n]}} \\
&= \frac{|\beta_0| \sum_{m=1}^{M} \sqrt{\alpha^{\varkappa}_m[n]}}{\sqrt{d^{\alpha}_{Ip_k}[n] d^{\alpha}_{BI}[n]}}.
\end{aligned} \tag{15}$$

Let $\rho^{\varkappa}[n] = |\beta_0|^2(\sum_{m=1}^{M} \sqrt{\alpha^{\varkappa}_m[n]})^2$, all the channel power gain are represented as follows:

$$|\mathbf{v}^H_{p_k}[n]\mathbf{\Theta}^{\varkappa}[n]\mathbf{g}[n]|^2 \leqslant \frac{\rho^{\varkappa}[n]}{d^{\alpha}_{Ip_k}[n] d^{\alpha}_{BI}[n]}, \tag{16}$$

$$|\mathbf{v}^H_{p_k}[n]\mathbf{\Theta}^{\varkappa}[n]\mathbf{v}_{s_1}[n]|^2 \leqslant \frac{\rho^{\varkappa}[n]}{d^{\alpha}_{Ip_k}[n] d^{\alpha}_{Is_1}[n]}, \tag{17}$$

$$|\mathbf{v}^H_{s_2}[n]\mathbf{\Theta}^{\varkappa}[n]\mathbf{v}_{s_1}[n]|^2 \leqslant \frac{\rho^{\varkappa}[n]}{d^{\alpha}_{Is_2}[n] d^{\alpha}_{Is_1}[n]}, \tag{18}$$

$$|\mathbf{v}^H_{s_2}[n]\mathbf{\Theta}^{\varkappa}[n]\mathbf{g}[n]|^2 \leqslant \frac{\rho^{\varkappa}[n]}{d^{\alpha}_{Is_2}[n] d^{\alpha}_{BI}[n]}. \tag{19}$$

Based on (16)-(19), the rate function of PUs as $R^{\mathbf{Q}}_{p_k}[n]$, and the achievable rate function of $s_2$ as $R^{\mathbf{Q}}_s[n]$, are proposed under the trajectory design of STAR-RIS-assisted UAV for time slot $n$ as follows:

$$R^{\mathbf{Q}}_{p_k}[n] = \log_2\left(1 + \frac{p_{p_k}[n]\rho^{\varkappa}[n]}{\breve{d}^{\frac{\alpha}{2}}_{Ip_k}[n]\breve{d}^{\frac{\alpha}{2}}_{BI}[n]\chi_{p_k}[n]}\right), \tag{20}$$

$$R^{\mathbf{Q}}_{s}[n] = \log_2\left(1 + \frac{p_{s_2}[n]\rho^{\varkappa}[n]}{\breve{d}^{\frac{\alpha}{2}}_{Is_2}[n]\breve{d}^{\frac{\alpha}{2}}_{Is_1}[n]\chi_{s}[n]}\right), \tag{21}$$

where the introduced sets $\mathbf{X} = \{\chi_{p_k}[n], \chi_s[n] \big| \; \forall\, k \in \mathcal{K}, n = 1, \ldots, N-1\}$, and $\mathbf{D} = \{\breve{d}_{BI}[n],\; \breve{d}_{Iu}[n] \big| \; \forall u \in \mathcal{U},\; n = 1, \ldots, N-1\}$ are contains of auxiliary variables. As a result, we can

reformulate the optimization problem (13) as:

$$\max_{\mathbf{Q},\mathbf{X},\mathbf{D}} \sum_{n=1}^{N-1}\left(\sum_{k=1}^{K} R_{p_k}^{\mathbf{Q}}[n] + R_s^{\mathbf{Q}}[n]\right) \tag{22a}$$

$$\text{s.t. } \|\mathbf{q}[n]-\mathbf{q}[n-1]\| \leqslant V_{\max}\delta,\ n=1,\ldots,N, \tag{22b}$$

$$R_{p_k}^{\mathbf{Q}}[n] \geqslant R^{\text{rsv}},\ \forall k\in\mathcal{K},\ n=1,\ldots,N-1, \tag{22c}$$

$$\frac{(\sum_{w\in\xi[n]} p_{p_w}[n])\rho^{\varkappa}[n]}{\breve{d}_{Ip_k}^{\frac{\alpha}{2}}[n]\breve{d}_{BI}^{\frac{\alpha}{2}}[n]} + \frac{p_{s_2}[n]\rho^{\varkappa}[n]}{\breve{d}_{Ip_k}^{\frac{\alpha}{2}}[n]\breve{d}_{Is_1}^{\frac{\alpha}{2}}[n]} + \sigma^2 \leqslant \chi_{p_k}[n],$$
$$\varkappa\in\Xi,\ \forall k\in\mathcal{K},\ n=1,\ldots,N-1, \tag{22d}$$

$$\frac{(\sum_{k=1}^{K} p_{p_k}[n])\rho^{\varkappa}[n]}{\breve{d}_{Is_2}^{\frac{\alpha}{2}}[n]\breve{d}_{BI}^{\frac{\alpha}{2}}[n]} + \sigma^2 \leqslant \chi_s[n],\ \varkappa\in\Xi, n=1,\ldots,N-1, \tag{22e}$$

$$\|\mathbf{q}[n]-\mathbf{w}_B\|^2 + H^2 \leqslant \breve{d}_{BI}[n],\ n=1,\ldots,N-1, \tag{22f}$$

$$\|\mathbf{q}[n]-\mathbf{r}_u\|^2 + H^2 \leqslant \breve{d}_{Iu}[n],\ \forall u\in\mathcal{U},\ n=1,\ldots,N-1. \tag{22g}$$

Although, the optimization problem (22) is still non-convex. To tackle the non-convexity of the STAR-RIS-assisted UAV trajectory design problem, GP as a fast and easy converting method to the convex optimization problem is obtained [28]. Hence, (22b) is rewritten as follows:

$$\sqrt{\left(x_q[n]-x_q[n-1]\right)^2 + \left(y_q[n]-y_q[n-1]\right)^2} \leqslant V_{\max}\delta, \tag{23}$$

then, we present the fractional upper bound inequality of two consecutive points as follows:

$$\frac{x_q^2[n]+x_q^2[n-1]+y_q^2[n]+y_q^2[n-1]}{2x_q[n]x_q[n-1]+2y_q[n]y_q[n-1]+(V_{\max}\delta)^2} \leqslant 1. \tag{24}$$

Subsequently, to obtain the GP standard form and achieve a convex representation of (24), the non-posynomial function is transformed into posynomial form by applying the arithmetic-geometric mean approximation (AGMA) [29], as detailed below:

$$\left(x_q^2[n]+x_q^2[n-1]+y_q^2[n]+y_q^2[n-1]\right) \times\left(\frac{2x_q[n]x_q[n-1]}{\eta^1[n]}\right)^{-\eta^1[n]} \times\left(\frac{2y_q[n]y_q[n-1]}{\eta^2[n]}\right)^{-\eta^2[n]} \times\left(\frac{(V_{\max}\delta)^2}{\eta^3[n]}\right)^{-\eta^3[n]} \leqslant 1,\ n=1,\ldots,N, \tag{25}$$

where

$$\eta^1[n] = \frac{2x_q^\tau[n]x_q^\tau[n-1]}{2x_q^\tau[n]x_q^\tau[n-1]+2y_q^\tau[n]y_q^\tau[n-1]+(V_{\max}\delta)^2}, \tag{26}$$

$$\eta^2[n] = \frac{2y_q^\tau[n]y_q^\tau[n-1]}{2x_q^\tau[n]x_q^\tau[n-1]+2y_q^\tau[n]y_q^\tau[n-1]+(V_{\max}\delta)^2}, \tag{27}$$

$$\eta^3[n] = \frac{(V_{\max}\delta)^2}{2x_q^\tau[n]x_q^\tau[n-1]+2y_q^\tau[n]y_q^\tau[n-1]+(V_{\max}\delta)^2}, \tag{28}$$

and $\tau$ is the iteration index, which refers to the previous step. Next, to facilitate convert the objective function (22a) and the constraint (22c) in the convex form, two new variables $\Upsilon_{p_k}[n]$, and $\Upsilon_s[n]$ are defined as:

$$\Upsilon_{p_k}[n] = \frac{\breve{d}_{Ip_k}^{\frac{\alpha}{2}}[n]\breve{d}_{BI}^{\frac{\alpha}{2}}[n]\chi_{p_k}[n]}{\breve{d}_{Ip_k}^{\frac{\alpha}{2}}[n]\breve{d}_{BI}^{\frac{\alpha}{2}}[n]\chi_{p_k}[n] + p_{p_k}[n]\rho^{\varkappa}[n]}, \tag{29}$$

$$\Upsilon_s[n] = \frac{\breve{d}_{Is_2}^{\frac{\alpha}{2}}[n]\breve{d}_{Is_1}^{\frac{\alpha}{2}}[n]\chi_s[n]}{\breve{d}_{Is_2}^{\frac{\alpha}{2}}[n]\breve{d}_{Is_1}^{\frac{\alpha}{2}}[n]\chi_s[n] + p_{s_2}[n]\rho^{\varkappa}[n]}. \tag{30}$$

Then, based on the GP form, $\Upsilon_{p_k}[n]$ can be approximated as follows:

$$\Upsilon_{p_k}[n] \approx \left(\breve{d}_{Ip_k}^{\frac{\alpha}{2}}[n]\breve{d}_{BI}^{\frac{\alpha}{2}}[n]\chi_{p_k}[n]\right)\times\left(\frac{\breve{d}_{Ip_k}^{\frac{\alpha}{2}}[n]\breve{d}_{BI}^{\frac{\alpha}{2}}[n]\chi_{p_k}[n]}{\kappa_{p_k}^1[n]}\right)^{-\kappa_{p_k}^1[n]} \times\left(\frac{p_{p_k}[n]\rho^{\varkappa}[n]}{\kappa_{p_k}^2[n]}\right)^{-\kappa_{p_k}^2[n]},\ \forall k\in\mathcal{K},\ n=1,\ldots,N-1, \tag{31}$$

where

$$\kappa_{p_k}^1[n] = \frac{\left(\breve{d}_{Ip_k}^\tau[n]\right)^{\frac{\alpha}{2}}\left(\breve{d}_{BI}^\tau[n]\right)^{\frac{\alpha}{2}}\chi_{p_k}^\tau[n]}{\left(\breve{d}_{Ip_k}^\tau[n]\right)^{\frac{\alpha}{2}}\left(\breve{d}_{BI}^\tau[n]\right)^{\frac{\alpha}{2}}\chi_{p_k}^\tau[n] + p_{p_k}[n]\rho^{\varkappa}[n]}, \tag{32}$$

$$\kappa_{p_k}^2[n] = \frac{p_{p_k}[n]\rho^{\varkappa}[n]}{\left(\breve{d}_{Ip_k}^\tau[n]\right)^{\frac{\alpha}{2}}\left(\breve{d}_{BI}^\tau[n]\right)^{\frac{\alpha}{2}}\chi_{p_k}^\tau[n] + p_{p_k}[n]\rho^{\varkappa}[n]}. \tag{33}$$

Similarly, the convex representation of $\Upsilon_s[n]$ derived from the GP approximation is proposed as follows:

$$\Upsilon_s[n] \approx \left(\breve{d}_{Is_2}^{\frac{\alpha}{2}}[n]\breve{d}_{Is_1}^{\frac{\alpha}{2}}[n]\chi_s[n]\right)\times\left(\frac{\breve{d}_{Is_2}^{\frac{\alpha}{2}}[n]\breve{d}_{Is_1}^{\frac{\alpha}{2}}[n]\chi_s[n]}{\kappa_s^1[n]}\right)^{-\kappa_s^1[n]} \times\left(\frac{p_{s_2}[n]\rho^{\varkappa}[n]}{\kappa_s^2[n]}\right)^{-\kappa_s^2[n]},\ n=1,\ldots,N-1, \tag{34}$$

where

$$\kappa_s^1[n] = \frac{\left(\breve{d}_{Is_2}^\tau[n]\right)^{\frac{\alpha}{2}}\left(\breve{d}_{Is_1}^\tau[n]\right)^{\frac{\alpha}{2}}\chi_s^\tau[n]}{\left(\breve{d}_{Is_2}^\tau[n]\right)^{\frac{\alpha}{2}}\left(\breve{d}_{Is_1}^\tau[n]\right)^{\frac{\alpha}{2}}\chi_s^\tau[n] + p_{s_2}[n]\rho^{\varkappa}[n]}, \tag{35}$$

$$\kappa_s^2[n] = \frac{p_{s_2}[n]\rho^{\varkappa}[n]}{\left(\breve{d}_{Is_2}^\tau[n]\right)^{\frac{\alpha}{2}}\left(\breve{d}_{Is_1}^\tau[n]\right)^{\frac{\alpha}{2}}\chi_s^\tau[n] + p_{s_2}[n]\rho^{\varkappa}[n]}. \tag{36}$$

Ultimately, the convex formulation of constraint (22c) is articulated as follows:

$$\Upsilon_{p_k}[n] \leqslant 2^{-R^{\text{rsv}}[n]},\ \forall k\in\mathcal{K},\ n=1,\cdots,N-1. \tag{37}$$

By reformulating constraints (22d) and (22e), we can demonstrate that both conform to the conventional structure of GP approximations, where posynomial functions are constrained to be less than or equal to 1:

$$\left(\frac{(\sum_{w\in\xi[n]} p_{p_w}[n])\rho^{\varkappa}[n]}{\breve{d}_{Ip_k}^{\frac{\alpha}{2}}[n]\breve{d}_{BI}^{\frac{\alpha}{2}}[n]} + \frac{p_{s_2}[n]\rho^{\varkappa}[n]}{\breve{d}_{Ip_k}^{\frac{\alpha}{2}}[n]\breve{d}_{Is_1}^{\frac{\alpha}{2}}[n]} + \sigma^2\right)\chi_{p_k}^{-1}[n] \leqslant 1,$$
$$\varkappa\in\Xi,\ \forall k\in\mathcal{K},\ n=1,\cdots,N-1, \tag{38}$$

$$\left(\frac{(\sum_{k=1}^{K} p_{p_k}[n])\rho^{\varkappa}[n]}{\breve{d}_{Is_2}^{\frac{\alpha}{2}}[n]\breve{d}_{BI}^{\frac{\alpha}{2}}[n]} + \sigma^2\right)\chi_s^{-1}[n] \leqslant 1, \varkappa\in\Xi,\ n=1,\cdots,N-1. \tag{39}$$

The conversion of constraint (22f) into GP form follows a process similar to that of constraint (22b), which is outlined as follows:

$$\left(x_q^2[n]+x_B^2+y_q^2[n]+y_B^2+H^2\right)\times\left(\frac{2x_q[n]x_B}{\vartheta_B^1[n]}\right)^{-\vartheta_B^1[n]}\times\left(\frac{2y_q[n]y_B}{\vartheta_B^2[n]}\right)^{-\vartheta_B^2[n]}\times\left(\frac{\breve{d}_{BI}[n]}{\vartheta_B^3[n]}\right)^{-\vartheta_B^3[n]}\leqslant 1,\ n=1,\dots,N-1, \quad (40)$$

where

$$\vartheta_B^1[n]=\frac{2x_q^\tau[n]x_B}{2x_q^\tau[n]x_B+2y_q^\tau[n]y_B+\breve{d}_{BI}^\tau[n]}, \quad (41)$$

$$\vartheta_B^2[n]=\frac{2y_q^\tau[n]y_B}{2x_q^\tau[n]x_B+2y_q^\tau[n]y_B+\breve{d}_{BI}^\tau[n]}, \quad (42)$$

$$\vartheta_B^3[n]=\frac{\breve{d}_{BI}^\tau[n]}{2x_q^\tau[n]x_B+2y_q^\tau[n]y_B+\breve{d}_{BI}^\tau[n]}. \quad (43)$$

As well as, the GP form of constraint (22g) is:

$$\left(x_q^2[n]+x_u^2+y_q^2[n]+y_u^2+H^2\right)\times\left(\frac{2x_q[n]x_u}{\vartheta_u^1[n]}\right)^{-\vartheta_u^1[n]}\times\left(\frac{2y_q[n]y_u}{\vartheta_u^2[n]}\right)^{-\vartheta_u^2[n]}\times\left(\frac{\breve{d}_{Iu}[n]}{\vartheta_u^3[n]}\right)^{-\vartheta_u^3[n]}\leqslant 1,\ \forall u\in\mathcal{U},\ n=1,\dots,N-1, \quad (44)$$

where

$$\vartheta_u^1[n]=\frac{2x_q^\tau[n]x_u}{2x_q^\tau[n]x_u+2y_q^\tau[n]y_u+\breve{d}_{Iu}^\tau[n]}, \quad (45)$$

$$\vartheta_u^2[n]=\frac{2y_q^\tau[n]y_u}{2x_q^\tau[n]x_u+2y_q^\tau[n]y_u+\breve{d}_{Iu}^\tau[n]}, \quad (46)$$

$$\vartheta_u^3[n]=\frac{\breve{d}_{Iu}^\tau[n]}{2x_q^\tau[n]x_u+2y_q^\tau[n]y_u+\breve{d}_{Iu}^\tau[n]}. \quad (47)$$

As a result, we derive the equivalent standard GP form for the trajectory design of the STAR-RIS-assisted UAV optimization problem (22):

$$\min_{\mathbf{Q},\mathbf{X},\mathbf{D}}\ \prod_{n=1}^{N-1}\left(\Upsilon_s[n]\left(\prod_{k=1}^{K}\Upsilon_{p_k}[n]\right)\right), \quad (48a)$$

$$\text{s.t. } (25),\ (37),\ (38),\ (39),\ (40),\ (44). \quad (48b)$$

which can be solved using existing software toolboxes such as CVX [30].

### B. Beamforming Optimization

For any given STAR-RIS-assisted UAV trajectory design, and corresponding power allocation parameters $\{\mathbf{Q},\mathbf{P}\}$, the elements' beamforming optimization are described in problem (22), can be achieved by:

$$\max_{\mathbf{\Phi}}\ \sum_{n=1}^{N}\left(\sum_{k=1}^{K}R_{p_k}[n]+R_s[n]\right) \quad (49a)$$

$$\text{s.t. } R_{p_k}[n]\geqslant R^{\text{rsv}},\ \forall k\in\mathcal{K},\ n=1,\dots,N, \quad (49b)$$

$$\alpha_m^{\varkappa}[n]\in[0,1],\ \forall\varkappa\in\Xi,\ \forall m\in\mathcal{M},\ n=1,\dots,N, \quad (49c)$$

$$\theta_m^{\varkappa}[n]\in[0,2\pi),\ \forall\varkappa\in\Xi,\ \forall m\in\mathcal{M},\ n=1,\dots,N, \quad (49d)$$

$$\alpha_m^t[n]+\alpha_m^r[n]=1,\ \forall m\in\mathcal{M},\ n=1,\dots,N, \quad (49e)$$

based on the objective function (49a) and the constraint (49b), the sub-problem (49) constitutes a non-convex maximization problem. Therefore, to achieve the optimal solution we conducted the process in two steps: phase shift optimization, and amplitude coefficient optimization.

*1) Phase shift optimization:* Due to reflections, diffraction, and scattering events, received signals through multiple paths may contain different delays and phase shifts. Since phases significantly influence the resultant amplitude, we implement phase alignment to achieve the maximum received energy [31]. Therefore, we establish the phase shift coefficients as:

$$\theta_m^{\varkappa}[n]=\begin{cases}\frac{2\pi d(m-1)}{\lambda}\left(\omega_{I\acute{u}}[n]-\omega_{BI}[n]\right)-\varpi_{\acute{u}}^{\varkappa}[n],\\ \frac{2\pi d(m-1)}{\lambda}\left(\omega_{I\acute{u}}[n]-\omega_{s_1I}[n]\right)-\varpi_{\acute{u}}^{\varkappa}[n],\end{cases} \quad (50)$$

where $\forall\varkappa\in\Xi,\ \forall m\in\mathcal{M},\ n=1,\dots,N,\ \acute{u}=\{p_k,s_2|\ k\in\mathcal{K}\}$, and $\varpi_u^{\varkappa}[n]\in[0,2\pi)$.

*2) Amplitude coefficient optimization:* In general, an efficient method for obtaining the optimal solution does not exist. Consequently, we employ a successive convex optimization technique to achieve sub-optimum solutions through an alternating approach. Hence, we denote the new definition for channel gains $\tilde{\mathbf{h}}_{BI\acute{u}}[n]=\text{diag}\left(\mathbf{v}_{\acute{u}}^H[n]\right)\mathbf{g}[n]$, and $\tilde{\mathbf{h}}_{s_1I\acute{u}}[n]=\text{diag}\left(\mathbf{v}_{\acute{u}}^H[n]\right)\mathbf{v}_{s_1}[n]$. Additionally, we define $\tilde{\mathbf{H}}_{BI\acute{u}}[n]=\tilde{\mathbf{h}}_{BI\acute{u}}[n]\tilde{\mathbf{h}}_{BI\acute{u}}^H[n]$, $\tilde{\mathbf{H}}_{s_1I\acute{u}}[n]=\tilde{\mathbf{h}}_{s_1I\acute{u}}[n]\tilde{\mathbf{h}}_{s_1I\acute{u}}^H[n]$, and $\boldsymbol{\Phi}^{\varkappa}[n]=\boldsymbol{\varphi}^{\varkappa}[n]\boldsymbol{\varphi}^{\varkappa H}[n]$ where $\boldsymbol{\Phi}^{\varkappa}[n]\succeq 0$, $\text{rank}(\boldsymbol{\Phi}^{\varkappa}[n])=1$, and $\boldsymbol{\Phi}^{\varkappa}{}_{m,m}[n]=\alpha_m^{\varkappa}[n],\ \forall\varkappa\in\Xi$. Therefore, we obtain $\left|\mathbf{v}_{\acute{u}}^H[n]\mathbf{\Theta}^{\varkappa}[n]\mathbf{g}[n]\right|^2=\left|\boldsymbol{\varphi}^{\varkappa H}[n]\tilde{\mathbf{h}}_{BI\acute{u}}[n]\right|^2=\text{Tr}\left(\boldsymbol{\Phi}^{\varkappa}[n]\tilde{\mathbf{H}}_{BI\acute{u}}[n]\right)$, and $\left|\mathbf{v}_{\acute{u}}^H[n]\mathbf{\Theta}^{\varkappa}[n]\mathbf{v}_{s_1}[n]\right|^2=\text{Tr}\left(\boldsymbol{\Phi}^{\varkappa}[n]\tilde{\mathbf{H}}_{s_1I\acute{u}}[n]\right)$.

On the other hand, for the sake of convenience, we introduce two slack variable sets, $\mathbf{E}=\{e_{p_k}[n],e_s[n]|\ k\in\mathcal{K},\ n=1,\dots,N\}$ and $\mathbf{F}=\{f_{p_k}[n],f_s[n]|\ k\in\mathcal{K},\ n=1,\dots,N\}$, which are constrained as (51)-(54). Consequently, the achievable rates $R_{p_k}^{\mathbf{\Phi}}[n]$ and $R_s^{\mathbf{\Phi}}[n]$ in beamforming optimization can be expressed as follows:

$$R_{p_k}^{\mathbf{\Phi}}[n]=\log_2\left(1+\frac{1}{e_{p_k}[n]f_{p_k}[n]}\right), \quad (55)$$

$$R_s^{\mathbf{\Phi}}[n]=\log_2\left(1+\frac{1}{e_s[n]f_s[n]}\right). \quad (56)$$

It can be observed that the achievable rate $\log_2\left(1+\frac{1}{e_{p_k}[n]f_{p_k}[n]}\right)$ is a joint convex function with respect to $e_{p_k}[n]$ and $f_{p_k}[n]$, under the positive definite condition of the Hessian matrix [32].

As denoted in [32], the first-order Taylor expansion of any concave function at a given point provides a global upper bound. Accordingly, we can derive the following

$$\frac{1}{e_{p_k}[n]} \leqslant p_{p_k}[n]\mathrm{Tr}\left(\boldsymbol{\Phi}^{\varkappa}[n]\tilde{\mathbf{H}}_{BIp_k}[n]\right),\ \varkappa \in \Xi,\ \forall k \in \mathcal{K},\ n = 1,\ldots,N, \tag{51}$$

$$f_{p_k}[n] \geqslant \sum_{w\in\xi[n]} p_{p_w}[n]\mathrm{Tr}\left(\boldsymbol{\Phi}^{\varkappa}[n]\tilde{\mathbf{H}}_{BIp_k}[n]\right) + p_{s_2}[n]\mathrm{Tr}\left(\boldsymbol{\Phi}^{\varkappa}[n]\tilde{\mathbf{H}}_{s_1Ip_k}[n]\right) + \sigma^2,\ \varkappa \in \Xi,\ \forall k \in \mathcal{K},\ n = 1,\ldots,N, \tag{52}$$

$$\frac{1}{e_s[n]} \leqslant p_{s_2}[n]\mathrm{Tr}\left(\boldsymbol{\Phi}^{\varkappa}[n]\tilde{\mathbf{H}}_{s_1Is_2}[n]\right),\ \varkappa \in \Xi,\ n = 1,\ldots,N, \tag{53}$$

$$f_s[n] \geqslant \sum_{k=1}^{K} p_{p_k}[n]\mathrm{Tr}\left(\boldsymbol{\Phi}^{\varkappa}[n]\tilde{\mathbf{H}}_{BIs_2}[n]\right) + \sigma^2,\ \varkappa \in \Xi,\ n = 1,\ldots,N, \tag{54}$$

---

$$R_{p_k}^{\mathbf{\Phi}}[n] \geqslant \log_2\left(1+\frac{1}{(e_{p_k}[n])^{\iota}(f_{p_k}[n])^{\iota}}\right) - \frac{(log_2 e)\left(e_{p_k}[n]-\left(e_{p_k}[n]\right)^{\iota}\right)}{\left(e_{p_k}[n]\right)^{\iota}\left(1+\left(\left(e_{p_k}[n]\right)^{\iota}\left(f_{p_k}[n]\right)^{\iota}\right)\right)} - \frac{(log_2 e)\left(f_{p_k}[n]-\left(f_{p_k}[n]\right)^{\iota}\right)}{\left(f_{p_k}[n]\right)^{\iota}\left(1+\left(\left(e_{p_k}[n]\right)^{\iota}\left(f_{p_k}[n]\right)^{\iota}\right)\right)} = \tilde{R}_{p_k}^{\mathbf{\Phi}}[n]. \tag{57}$$

Similarly,

$$R_{s}^{\mathbf{\Phi}}[n] \geqslant \log_2\left(1+\frac{1}{(e_{s}[n])^{\iota}(f_{s}[n])^{\iota}}\right) - \frac{(log_2 e)\left(e_{s}[n]-(e_{s}[n])^{\iota}\right)}{(e_{s}[n])^{\iota}\left(1+\left((e_{s}[n])^{\iota}(f_{s}[n])^{\iota}\right)\right)} - \frac{(log_2 e)\left(f_{s}[n]-(f_{s}[n])^{\iota}\right)}{(f_{s}[n])^{\iota}\left(1+\left((e_{s}[n])^{\iota}(f_{s}[n])^{\iota}\right)\right)} = \tilde{R}_{s}^{\mathbf{\Phi}}[n], \tag{58}$$

---

convex lower bound in each iteration $\iota$ at (57), and (58). Here, the iteration index of the algorithm is represented by $\iota$ which is used to obtain the STAR-RIS-assisted UAV's beamforming design based on the previous iteration step. With the aforementioned variable definitions, optimization problem (49) is reformulated as follows:

$$\max_{\mathbf{\Phi},\mathbf{E},\mathbf{F}} \sum_{n=1}^{N}\left(\sum_{k=1}^{K}\tilde{R}_{p_k}^{\mathbf{\Phi}}[n]+\tilde{R}_{s}^{\mathbf{\Phi}}[n]\right) \tag{59a}$$

$$\text{s.t. } \tilde{R}_{p_k}^{\mathbf{\Phi}}[n] \geqslant R^{\text{rsv}},\ \forall k \in \mathcal{K},\ n = 1,\ldots,N, \tag{59b}$$

$$0 \leqslant \alpha_m^{\varkappa}[n] \leqslant 1,\ \forall \varkappa \in \Xi,\ \forall m \in \mathcal{M},\ n = 1,\ldots,N, \tag{59c}$$

$$\alpha_m^{t}[n] + \alpha_m^{r}[n] = 1,\ \forall m \in \mathcal{M},\ n = 1,\ldots,N, \tag{59d}$$

$$\boldsymbol{\Phi}^{\varkappa}[n] \succeq 0,\ \forall \varkappa \in \Xi,\ n = 1,\ldots,N, \tag{59e}$$

$$\mathrm{rank}\left(\boldsymbol{\Phi}^{\varkappa}[n]\right) = 1,\ \forall \varkappa \in \Xi,\ n = 1,\ldots,N, \tag{59f}$$

$$\boldsymbol{\Phi}^{\varkappa}{}_{m,m}[n] = \alpha_m^{\varkappa}[n],\ \forall \varkappa \in \Xi,\ \forall m \in \mathcal{M},\ n = 1,\ldots,N, \tag{59g}$$

$$(51)-(54). \tag{59h}$$

It is important to observe that the persistence of non-convexity in problem (59) is attributable to the rank-one constraint presented in (59f). Particularly, we express the identical form of the $\mathrm{rank}\left(\boldsymbol{\Phi}^{\varkappa}[n]\right) = 1$ as $\lambda_{max}\left(\boldsymbol{\Phi}^{\varkappa}[n]\right) = \mathrm{Tr}\left(\boldsymbol{\Phi}^{\varkappa}[n]\right)$, where $\lambda_{max}\left(\boldsymbol{\Phi}^{\varkappa}[n]\right)$ is the largest eigenvalue of matrix $\boldsymbol{\Phi}^{\varkappa}[n]$ in time slot $n$. In addition, $\lambda_{max}\left(\boldsymbol{\Phi}^{\varkappa}[n]\right)$ can be represented as $\lambda_{max}\left(\boldsymbol{\Phi}^{\varkappa}[n]\right) = \boldsymbol{\nu}_{max}^{H}\left((\boldsymbol{\Phi}^{\varkappa}[n])^{\iota}\right)\boldsymbol{\Phi}^{\varkappa}[n]\boldsymbol{\nu}_{max}\left((\boldsymbol{\Phi}^{\varkappa}[n])^{\iota}\right)$ where $\boldsymbol{\nu}_{max}\left((\boldsymbol{\Phi}^{\varkappa}[n])^{\iota}\right)$ is eigenvector corresponding to the largest eigenvalue of matrix $\boldsymbol{\Phi}^{\varkappa}[n]$. As a consequence, to address the challenges posed by the non-convexity of constraint (59f), we implemented the sequential rank-one constraint relaxation algorithm as follows:

$$\lambda_{max}\left(\boldsymbol{\Phi}^{\varkappa}[n]\right) \geqslant (\zeta^{\varkappa})^{\iota}[n]\mathrm{Tr}\left(\boldsymbol{\Phi}^{\varkappa}[n]\right). \tag{60}$$

Here, $(\zeta^{\varkappa})^{\iota}[n] \in [0,1]$ serves as a relaxation parameter, regulating the ratio of the largest eigenvalue to the trace of matrix $\boldsymbol{\Phi}^{\varkappa}[n]$ in time slot $n$ and iteration $\iota$. Finally, the convex form of $\mathrm{rank}\left(\boldsymbol{\Phi}^{\varkappa}[n]\right) = 1$ is obtained as [33]:

$$\boldsymbol{\nu}_{max}^{H}\left((\boldsymbol{\Phi}^{\varkappa}[n])^{\iota}\right)\boldsymbol{\Phi}^{\varkappa}[n]\boldsymbol{\nu}_{max}\left((\boldsymbol{\Phi}^{\varkappa}[n])^{\iota}\right) \geqslant (\zeta^{\varkappa})^{\iota}[n]\mathrm{Tr}\left(\boldsymbol{\Phi}^{\varkappa}[n]\right),\quad \forall \varkappa \in \Xi,\ n = 1,\ldots,N, \tag{61}$$

further, the parameter $(\zeta^{\varkappa})^{\iota}[n]$ will undergo an update, with $(\delta^{\varkappa})^{\iota}[n]$ representing the step size during iteration $\iota$ at timeslot $n$:

$$(\zeta^{\varkappa})^{\iota}[n] = min\left(1, \frac{\lambda_{max}\left((\boldsymbol{\Phi}^{\varkappa}[n])^{\iota}\right)}{\mathrm{Tr}\left((\boldsymbol{\Phi}^{\varkappa}[n])^{\iota}\right)} + (\delta^{\varkappa})^{\iota}[n]\right),\quad \forall \varkappa \in \Xi,\ n = 1,\ldots,N. \tag{62}$$

Consequently, we present the convex formulation of the optimization problem (59) as follows:

$$\max_{\mathbf{\Phi},\mathbf{E},\mathbf{F}} \sum_{n=1}^{N}\left(\sum_{k=1}^{K}\tilde{R}_{p_k}^{\mathbf{\Phi}}[n]+\tilde{R}_{s}^{\mathbf{\Phi}}[n]\right) \tag{63a}$$

$$\text{s.t.}\quad (51)-(54),\ (59b)-(59e),\ (59g),\ (61). \tag{63b}$$

The achieved problem formulation (63) represents a standard convex semidefinite programming (SDP) formulation, which is addressed efficiently using numerical solvers, such as the SDP mode available in the CVX tool [30].

### C. Power Allocation Optimization

The power allocation issue associated with a specified STAR-RIS-assisted UAV trajectory design $\mathbf{Q}$ and the corresponding beamforming optimization $\mathbf{\Phi}$ will be addressed

$$R^{\mathbf{P}}_{p_k}[n] = \log_2\left(1 + \frac{p_{p_k}[n]\left|h_{BIp_k}[n]\right|^2}{\sum_{w\in\xi[n]} p_{p_w}[n]\left|h_{BIp_k}[n]\right|^2 + p_{s_2}[n]\left|h_{s_1Ip_k}[n]\right|^2 + \sigma^2}\right) \geqslant \log_2\left(\sum_{w\in\varrho[n]} p_{p_w}[n]\left|h_{BIp_k}[n]\right|^2 + p_{s_2}[n]\left|h_{s_1Ip_k}[n]\right|^2 + \sigma^2\right)$$
$$- \log_2\left(\sum_{w\in\xi[n]} p^{\mu}_{p_w}[n]\left|h_{BIp_k}[n]\right|^2 + p_{s_2}[n]\left|h_{s_1Ip_k}[n]\right|^2 + \sigma^2\right) - \frac{(log_2 e)\left|h_{BIp_k}[n]\right|^2\left(\sum_{w\in\xi[n]}\left(p_{p_w}[n] - p^{\mu}_{p_w}[n]\right)\right)}{\sum_{w\in\xi[n]} p^{\mu}_{p_w}[n]\left|h_{BIp_k}[n]\right|^2 + p_{s_2}[n]\left|h_{s_1Ip_k}[n]\right|^2 + \sigma^2} = \tilde{R}^{\mathbf{P}}_{p_k}[n]. \tag{65}$$

where $\varrho[n] = \{|h_{BIp_w}[n]|^2 \geqslant |h_{BIp_k}[n]|^2 | \forall\ w,\ k \in \mathcal{K}\}$. Similarly:

$$R^{\mathbf{P}}_{s}[n] = \log_2\left(1 + \frac{p_{s_2}[n]\left|h_{s_1Is_2}[n]\right|^2}{\sum_{k=1}^{K} p_{p_k}[n]\left|h_{BIs_2}[n]\right|^2 + \sigma^2}\right) \geqslant \log_2\left(p_{s_2}[n]\left|h_{s_1Is_2}[n]\right|^2 + \sum_{k=1}^{K} p_{p_k}[n]\left|h_{BIs_2}[n]\right|^2 + \sigma^2\right)$$
$$- \log_2\left(\sum_{k=1}^{K} p^{\mu}_{p_k}[n]\left|h_{BIs_2}[n]\right|^2 + \sigma^2\right) - \frac{(log_2 e)\left|h_{BIs_2}[n]\right|^2\left(\sum_{k=1}^{K}\left(p_{p_k}[n] - p^{\mu}_{p_k}[n]\right)\right)}{\left(\sum_{k=1}^{K} p^{\mu}_{p_k}[n]\left|h_{BIs_2}[n]\right|^2 + \sigma^2\right)} = \tilde{R}^{\mathbf{P}}_{s}[n]. \tag{66}$$

by formulating the problem as follows:

$$\max_{\mathbf{P}} \sum_{n=1}^{N}\left(\sum_{k=1}^{K} R_{p_k}[n] + R_s[n]\right) \tag{64a}$$
$$\text{s.t.}\ \ R_{p_k}[n] \geqslant R^{\text{rsv}},\ \forall k \in \mathcal{K},\ n = 1, \ldots, N, \tag{64b}$$
$$\sum_{k=1}^{K} p_{p_k}[n] \leqslant P_{\max},\ n = 1, \ldots, N, \tag{64c}$$

where corresponding to the non-convex objective function (64a) and the non-convex constraint (64b), problem (64) becomes a non-convex optimization problem that is typically NP-hard and challenging to solve. Initially, we remember that $h_{BI\acute{u}}[n] = \mathbf{v}^H_{\acute{u}}[n]\mathbf{\Theta}^{\varkappa}[n]\mathbf{g}[n]$ and $h_{s_1I\acute{u}}[n] = \mathbf{v}^H_{\acute{u}}[n]\mathbf{\Theta}^{\varkappa}[n]\mathbf{v}_{s_1}[n]$. To overcome this challenge, we adopt the D.C. approximation. This approach ensures that the first-order Taylor expansion at any given point $\mu$ provides a global upper bound for any concave function [34]. Therefore, we can express the detailed of approximation in (65), (66). We propose the following approximation of problem formulation (64) in a convex form, which includes the lower bounds of $\tilde{R}^{\mathbf{P}}_{p_k}[n]$ and $\tilde{R}^{\mathbf{P}}_{s}[n]$:

$$\max_{\mathbf{P}} \sum_{n=1}^{N}\left(\sum_{k=1}^{K} \tilde{R}^{\mathbf{P}}_{p_k}[n] + \tilde{R}^{\mathbf{P}}_{s}[n]\right) \tag{67a}$$
$$\text{s.t.}\ \ \tilde{R}^{\mathbf{P}}_{p_k}[n] \geqslant R^{\text{rsv}},\ \forall k \in \mathcal{K},\ n = 1, \ldots, N, \tag{67b}$$
$$\sum_{k=1}^{K} p_{p_k}[n] \leqslant P_{\max},\ n = 1, \ldots, N. \tag{67c}$$

Convex optimization problem (67) can be effectively addressed with the established convex optimization solvers, including CVX [30].

## V. Simulation Results

The purpose of this section is to demonstrate the performance of using the STAR-RIS-assisted UAV with PD-NOMA technique in the CR network. BS is located at $[0, 500]^T$ m, while all PUs, and SU pairs are randomly distributed within a $1 \times 1$ km$^2$ area. The STAR-RIS-assisted UAV flies with maximum speed of $V_{\max} = 20$ m/s at fixed altitude $H = 80$ m during communication time $T = 60$ s. The number of PUs is assumed $K = 4$ with minimum rate requirement $R^{\text{rsv}} = 0.3$ bps/Hz. The first SU, $s_1$, serves the second SU, $s_2$, through STAR-RIS-assisted UAV with power transmission $p_{s_2}[n] = 0.2$ W, and the power budget is set to $P_{\max} = 25$ W. Other simulation parameters are set as: $M = 20$ elements, $N = 20$ time slots, $\sigma^2 = -174$ dB, $\alpha = 2.2$, and $\beta_0 = -30$ dB.

### A. Impact of Communication Time

The impact of communication time on the performance of the proposed system is illustrated in Fig.2. By increasing communication time $T$, the maximum distance flying limitation between two sequential points of STAR-RIS-assisted UAV is more loose. This allows the STAR-RIS-assisted UAV to fly closer to both PUs and SUs and stay more time in optimal locations to enhance spectrum-sharing efficiency in CR network. Therefore, by increasing communication time from $T = 55s$ to $T = 65s$, the sum rate increases 12%, too.

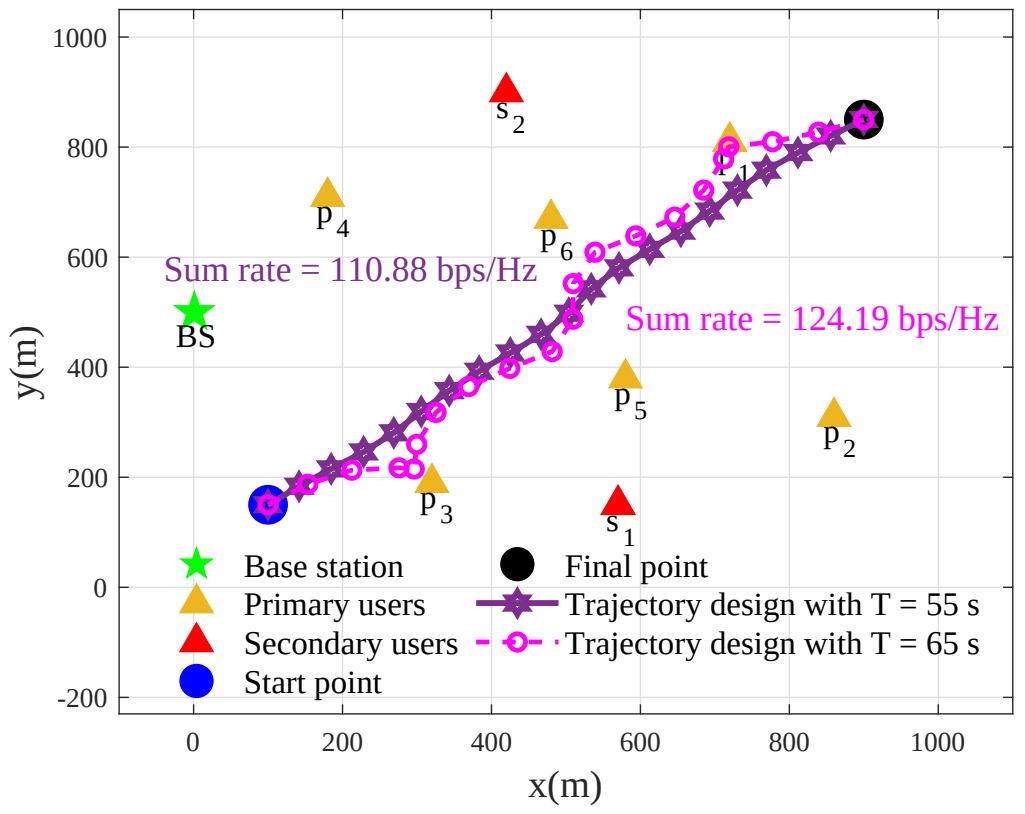

Fig. 2: Trajectory design of STAR-RIS-assisted UAV for $K = 6$, and different communication time $T$, in CR network.

### B. Impact of Users

This subsection studies how the number of users impacts system performance and the trajectory design of the STAR-RIS-assisted UAV in two fixed and mobile user scenarios.

*1) Number of PUs:* Fig. 3 presents the sum rate versus the number of PUs for different level of maximum power budget, $P_{\max}$. As demonstrated, the sum rate of the designed CR network grows with the increase of the PU's number, which leads to improving the overall capacity. On the other hand, the achievable sum rate of the STAR-RIS-assisted UAV under ES protocol with the PD-NOMA technique is increasing as $P_{\max}$ increases through the properties of interference management and SIC ordering exploitation.

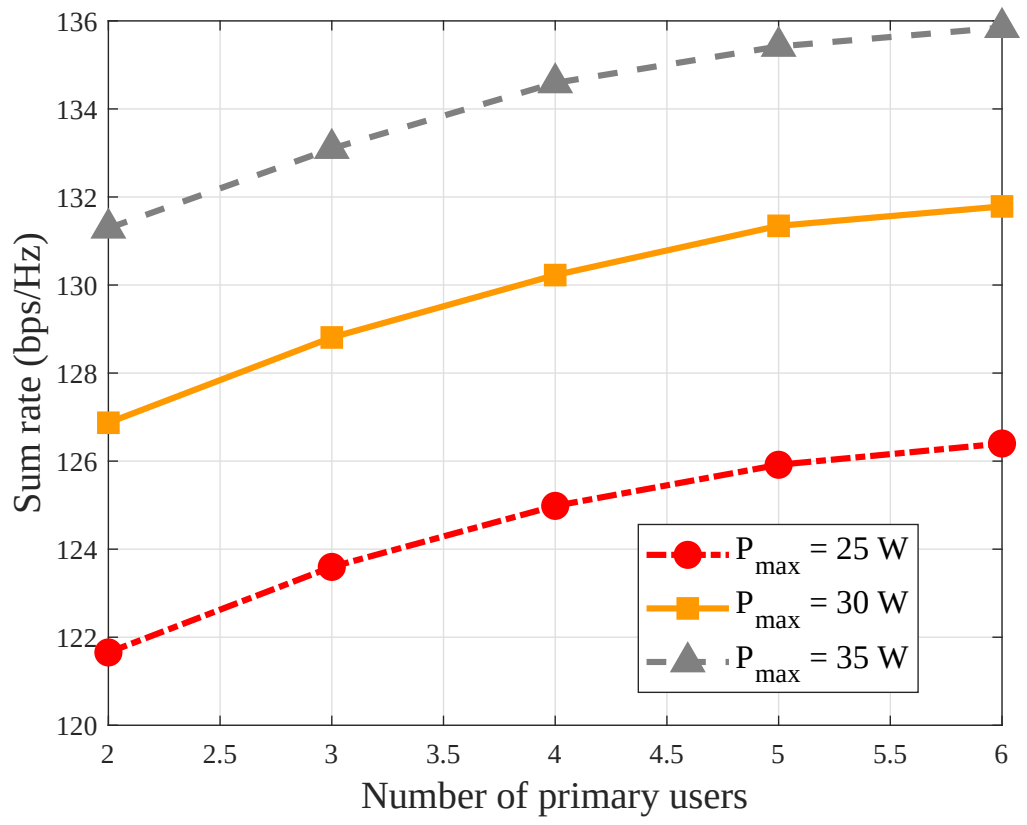

Fig. 3: Sum rate versus number of PUs for three cases: $P_{\max} = \{25, 30, 35\}$W.

*2) Number of SU Pairs:* As evaluated in Fig. 4, increasing the number of SU pairs leads to a decrease in the sum rate. This is because increasing SU pairs imposes higher sum interference to both the rate of PUs and SU pairs. In addition, due to STAR-RIS's potential to efficiently mitigate the interference of other users, we present the sum interference curve from SU pairs, which is growing weakly compared to the linear case as the number of SU pairs increases. Consequently, the STAR-RIS-assisted UAV not only improves the spectral efficiency and extends coverage in CR network but also improves the sum interference caused by increasing the SU pairs compared to traditional relays, which potentially amplifies interference.

*3) Mobility of PUs, and SU pairs:* To study the deployment of a practical scenario in a dynamic CR network, we investigate how the mobility of PUs and SU pairs affects the trajectory design of the STAR-RIS-assisted UAV with the PD-NOMA technique [35]. Fig. 5, shows the trajectory design of STAR-RIS-assisted UAV in the presence of six mobile PUs and a pair of mobile SUs. To maximize the sum rate of the proposed system model, the UAV's path, and incident signals to STAR-RIS should be designed, as well as resource allocation from the start point to the final point, during communication time $T = 65$ s.

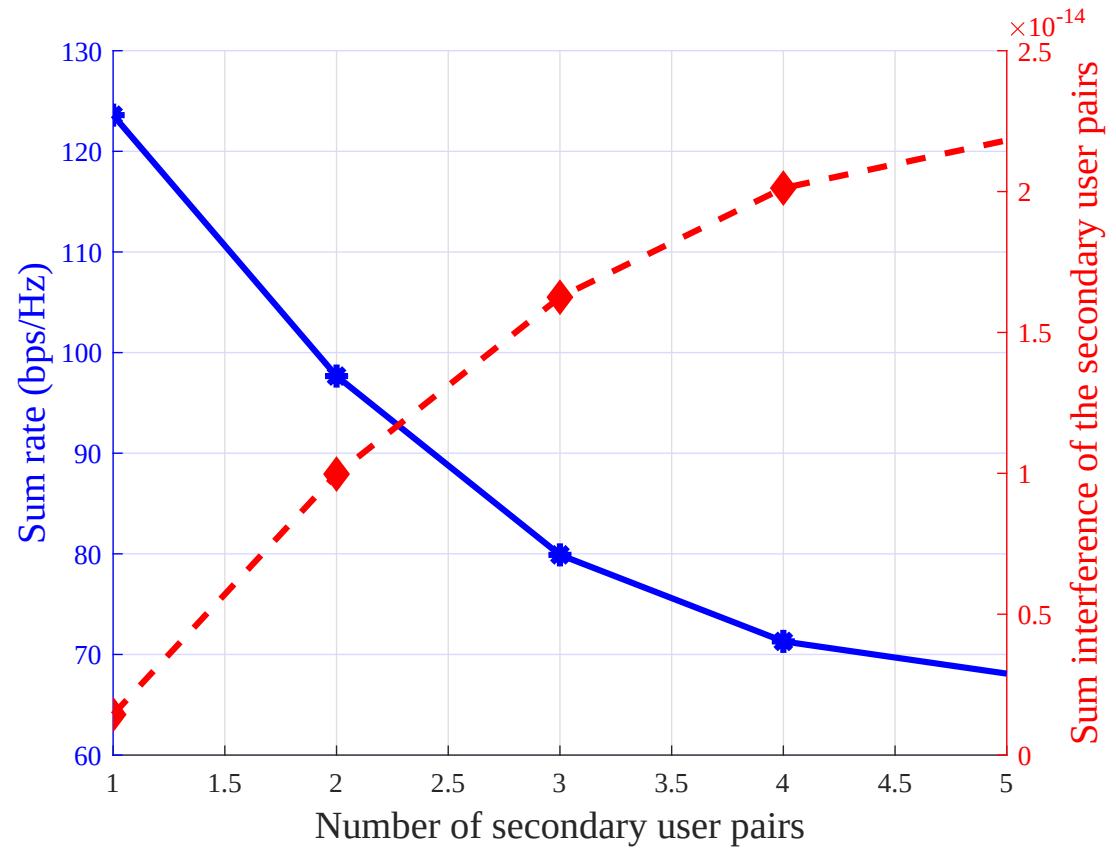

Fig. 4: Sum rate, and sum interference of the SU pairs versus the number of SU pairs, for $K = 3$.

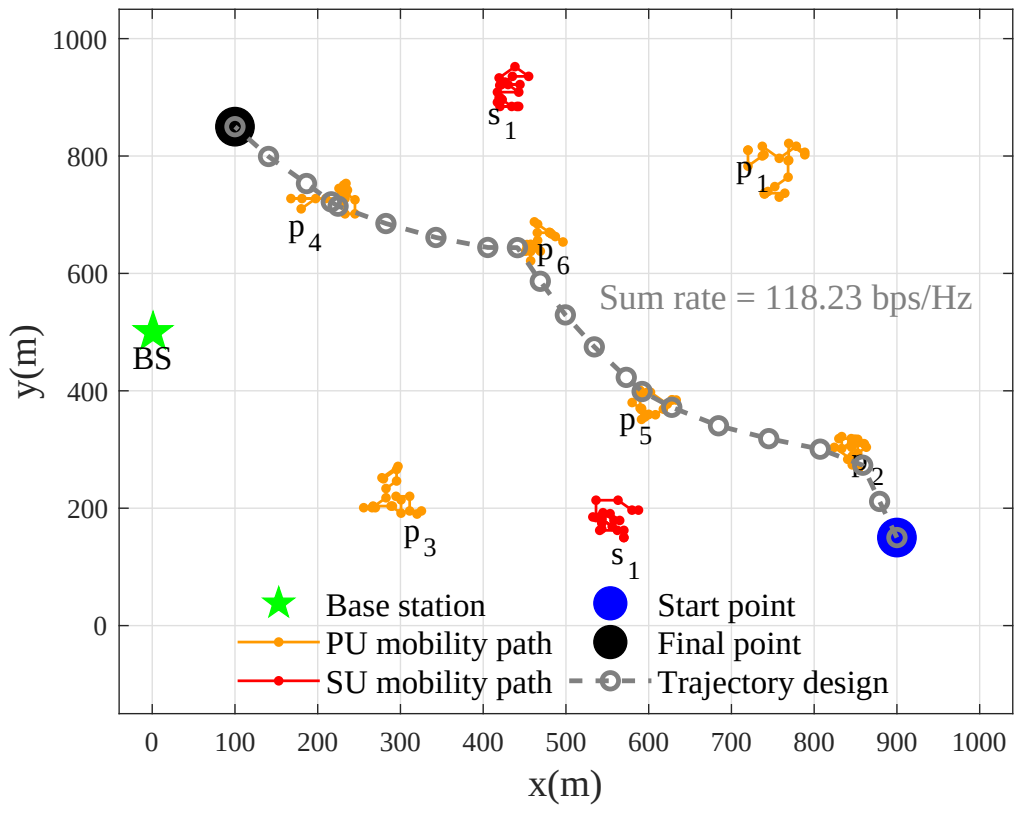

Fig. 5: Trajectory design of STAR-RIS-assisted UAV with mobility of six PUs and mobility of a SUs pair, for $T = 65$s.

Similar to Fig.5, Fig.6 gives a demonstration of the trajectory design of the STAR-RIS-assisted UAV with five SU pairs. The STAR-RIS effectively optimizes resource utilization by managing the interference caused by the SU pairs, enabling the UAV to design its path successfully to maximize the sum rate for both PUs and SU pairs under PUs' QoS.

### C. Impact of Elements

The illustration of Fig.7 shows the sum rate versus the number of elements for three intelligence surface modes, STAR-RIS mode, RIS mode (surface with reflecting elements only), and ITS mode (surface with transmitting elements only). In all three modes, expanding the number of intelligence surface elements leads to achieving a powerful sum rate. This increasing rate arises from the intelligence surface-assisted UAV with more elements, which enhances spatial diversity and addresses multipath propagation challenges. It enables multiple signal paths, provides greater control over incident signals, and optimizes resource allocation, to improve spectrum-sharing efficiency in the CR network for both PUs and SU pairs. It is obvious that the proposed STAR-RIS mode, which can simultaneously transmit and reflect the incident signals to both PUs and SUs on different sides of the intelligence surface, achieves a higher sum rate compared to the other two modes. On the other hand, the performance gap between ITS mode and RIS mode is related to the random distribution of the SU pair, which are positioned on both sides of the intelligence surface-assisted UAV. Therefore, SU pair connections are established only in ITS mode, resulting in a higher rate for that mode.

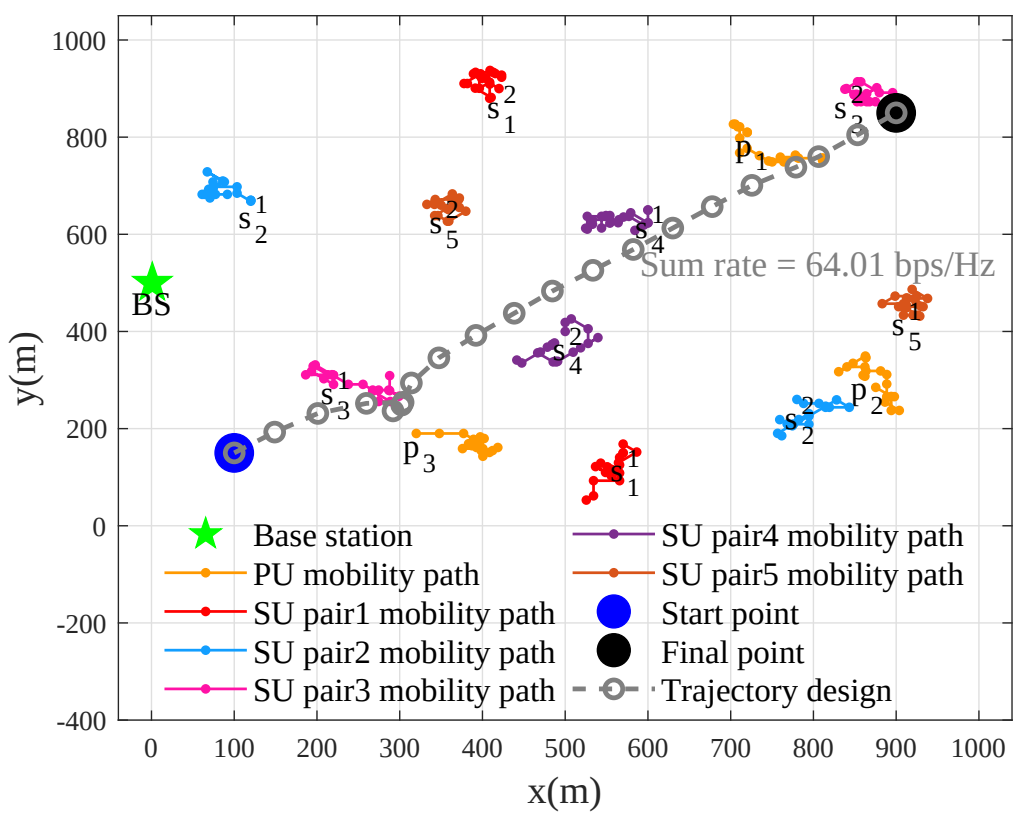

Fig. 6: Trajectory design of STAR-RIS-assisted UAV with mobility of three PUs and mobility of five SUs pairs, for $T = 65s$.

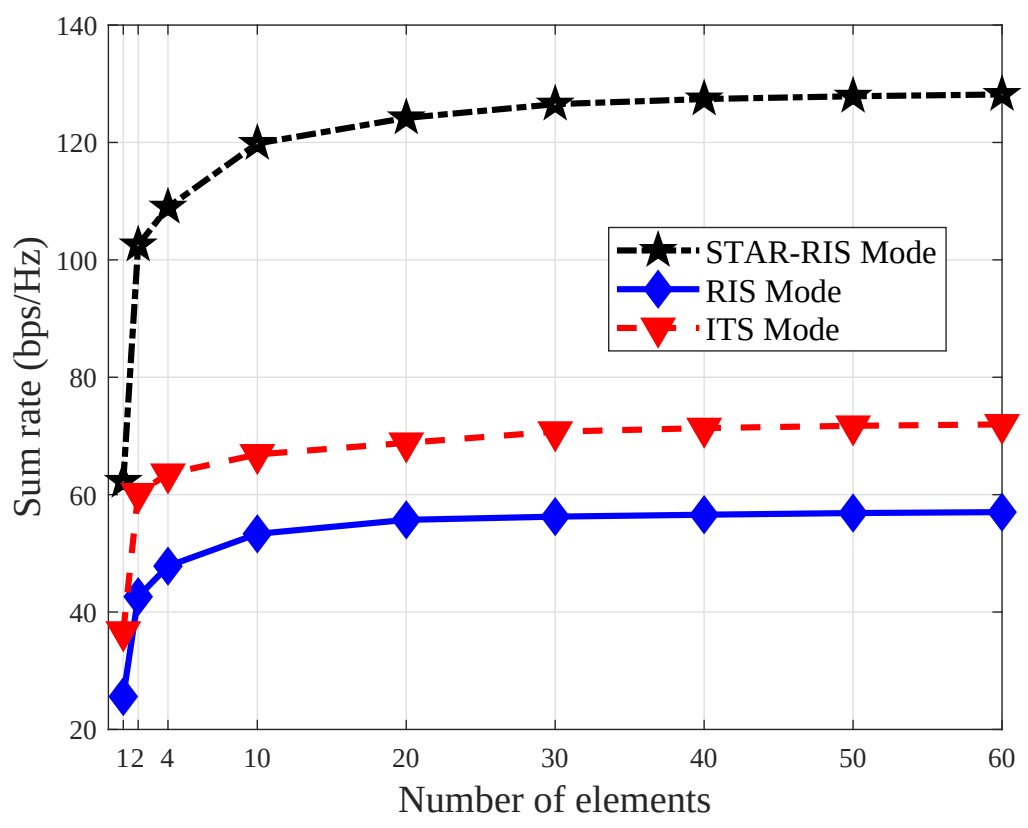

Fig. 7: Sum rate versus number of elements for three intelligence surface modes: STAR-RIS, RIS, ITS.

In Fig. 8, the innovation of the proposed system model is represented. Regardless of the STAR-RIS in a fixed environment, which determines the energy splitting ratio of elements due to the user distribution at the beginning of the communication and keeps them steady, the proposed scenario requires continuous adjustment during flight time. This arises from the dynamic channel gain between the STAR-RIS-assisted UAV and both PUs and SUs. Accordingly, we present the reflection amplitude and transmission amplitude versus the number of elements, and the time slot's number. In addition, the performance of the selected ES protocol in the proposed system model is satisfied by the constraint (12f). Therefore, Fig. 8a complements Fig.8b, and their amplitude summation equals 1.

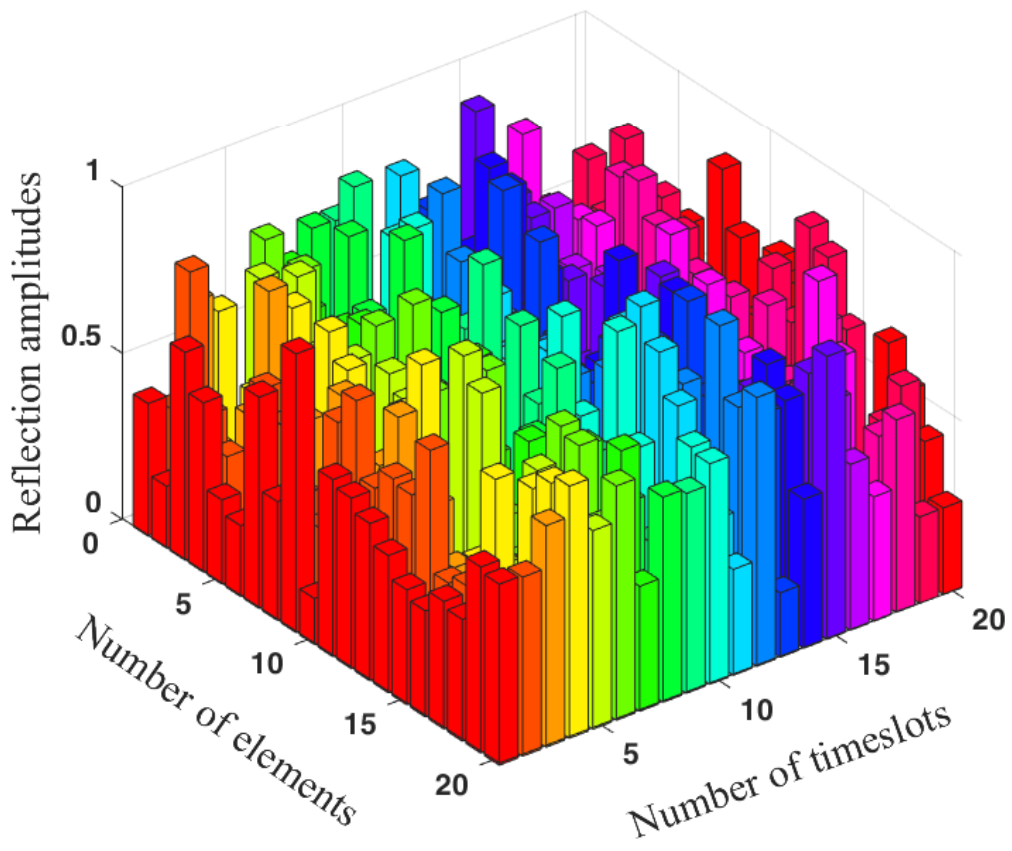

(a) The reflection amplitudes

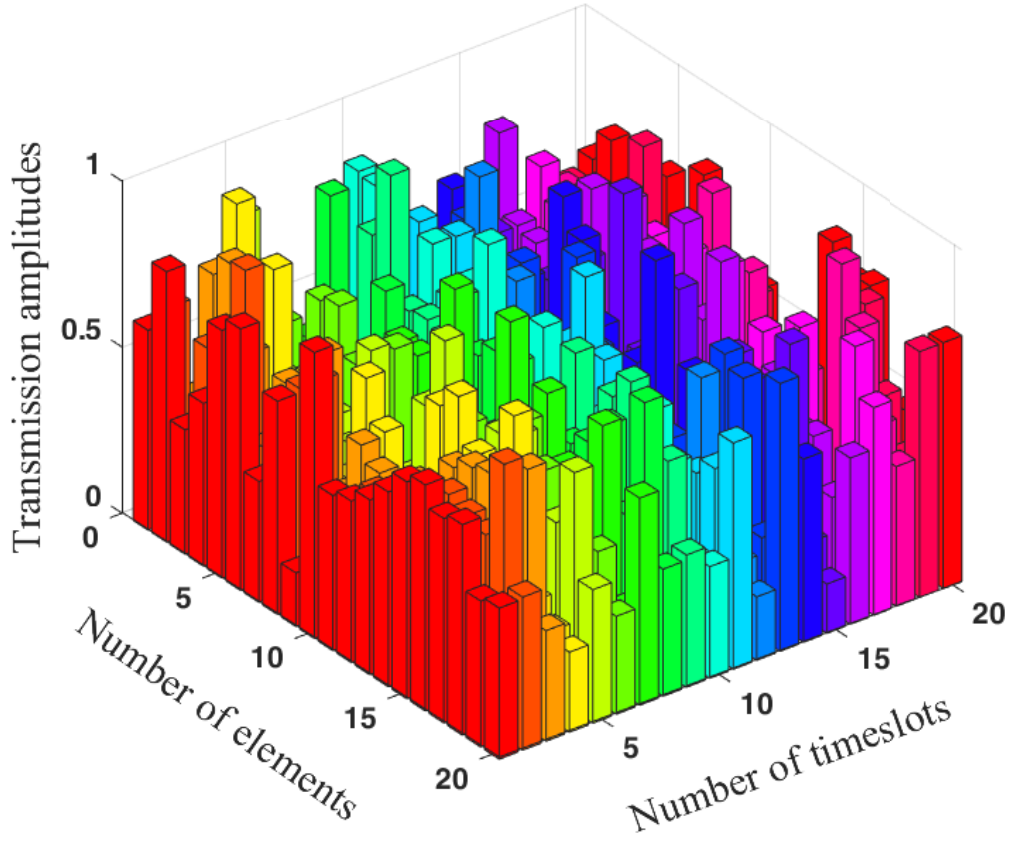

(b) The transmission amplitudes

Fig. 8: The energy splitting ratio of elements versus the number of elements, and time slot's number.

The sum rate versus the number of elements for different STAR-RIS operating protocols i.e., ES, MS, and TS is shown in Fig. 9. Similar to the conclusion of Fig. 7, by increasing the STAR-RIS's elements across three operating protocols, the sum rate increases, too. Since ES can jointly optimize the incident signal's amplitude and phase shift coefficients into transmitting and reflecting mode, it outperforms in dynamic environments. Therefore, ES is preferable for sum rate maximization of PUs and SUs. The MS protocol obtains a lower sum rate by optimizing each element's operation in only transmitting or reflecting mode. However, the TS protocol periodically optimizes all elements' operation modes in transmitting or reflecting, which nulls some users in each time slot.

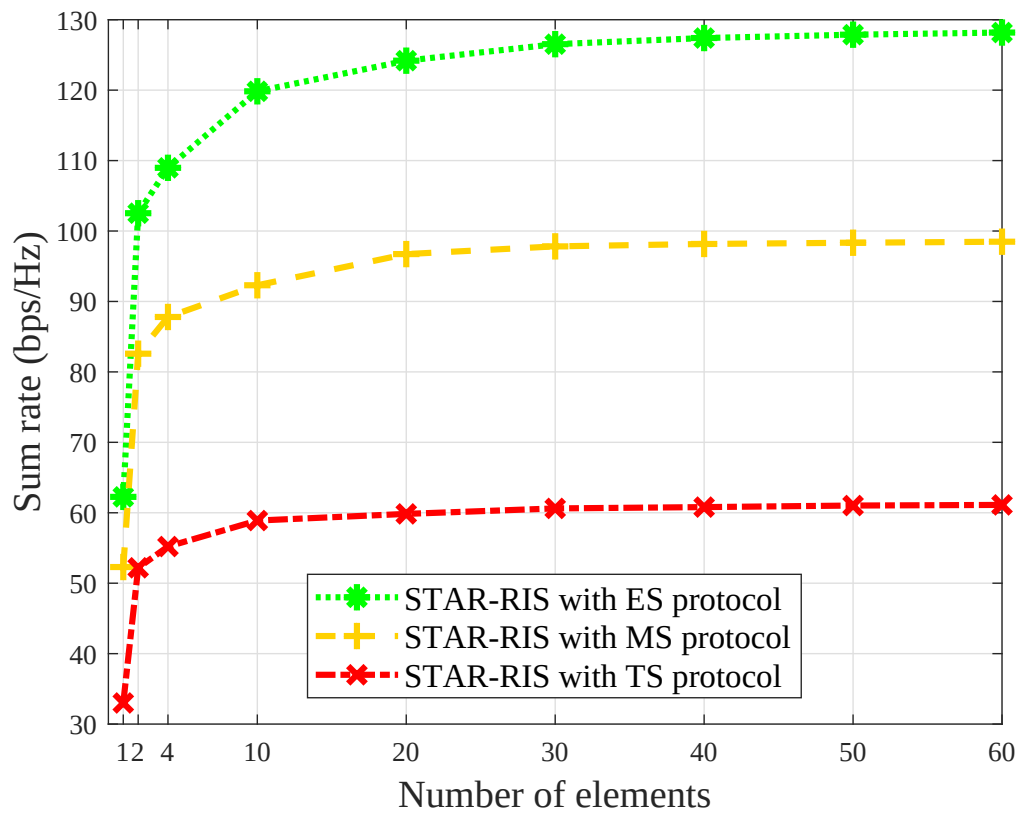

Fig. 9: Sum rate versus number of elements for three operating STAR-RIS protocols: ES, MS, and TS.

### D. Impact of Power Budget and Transmission Protocols

The sum rate versus the maximum power budget's variation is studied in Fig. 10. With increasing the power budget, the sum rate curves increase monotonically in all three intelligence surface modes: STAR-RIS, RIS, and ITS, due to enhanced users' access to resources. It is clear that the STAR-RIS-assisted UAV performs better than the traditional RIS/ITS-assisted UAV due to its ability to provide full coverage of both PUs and SUs during communication.

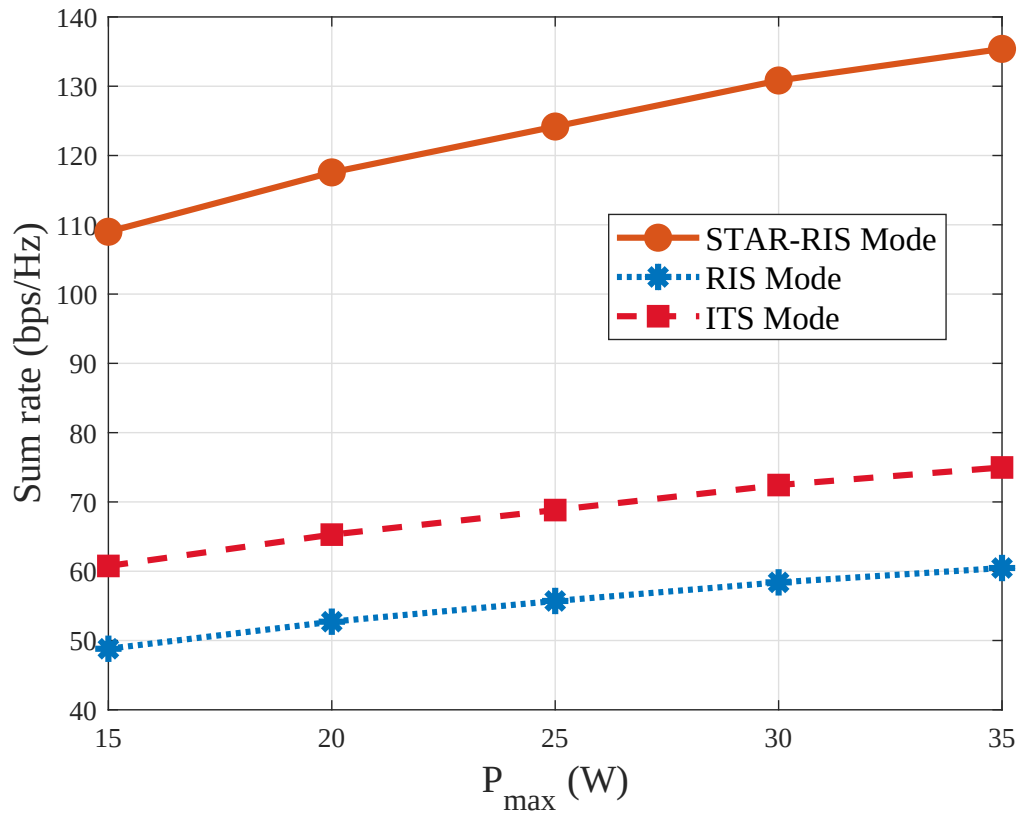

Fig. 10: Sum rate versus the maximum power budget for three intelligence surface modes: STAR-RIS, RIS, ITS.

Fig. 11 is presented, to examine the contribution of the transmission technique on the sum rate performance. It is worth noting that, the sum rate's gap between OMA and NOMA techniques is smaller with fewer time slots. However, this gap monotonically increases as the number of time slots increases. This is a result of STAR-RIS-assisted UAV with the PD-NOMA technique can serve both PUs and SUs with different channel conditions more effectively by sharing resources simultaneously compared to OMA, which divides resources orthogonally and may lead to some users being underserved and results in resource wastage. Consequently, the proposed system using the PD-NOMA technique improves 20% in performance achievement compared to the OMA technique.

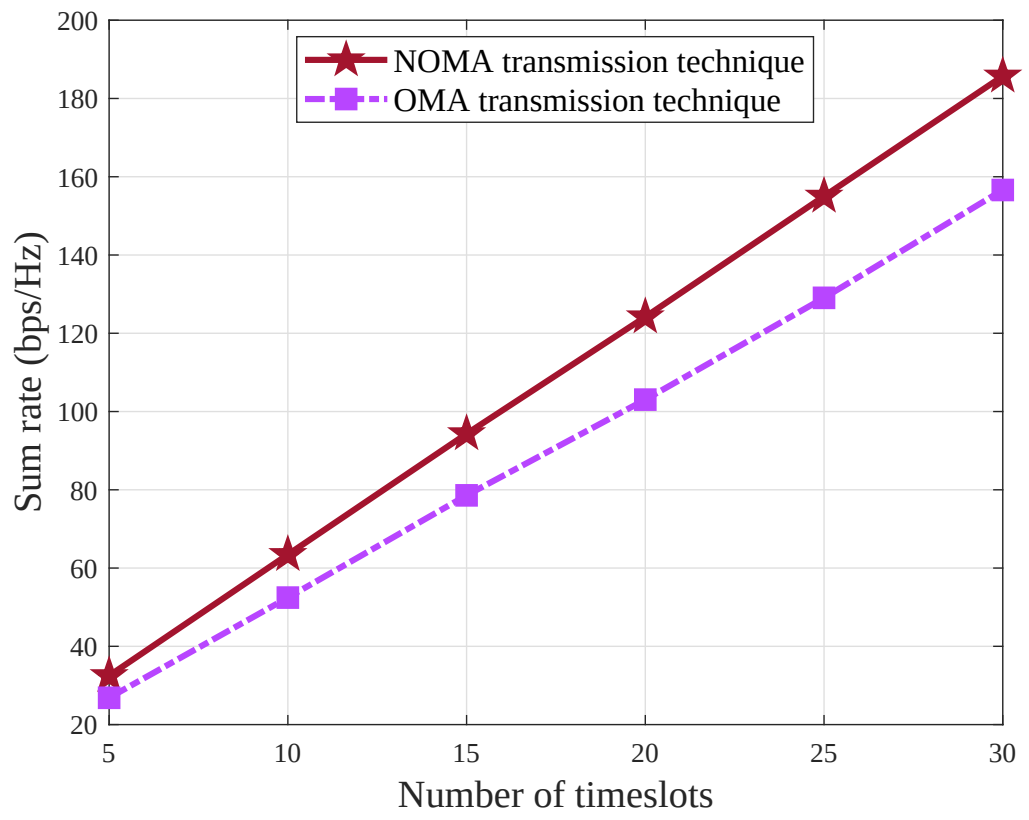

Fig. 11: Sum rate versus the time slot's number

.

## VI. Conclusion

This paper investigated a STAR-RIS-assisted UAV in CR network with spectrum-sharing approaches. In particular, the STAR-RIS-assisted UAV serves an area without any direct or significant links between BS and PUs, and BS to SUs pair. Hence, STAR-RIS-assisted UAV communicates with PUs using the PD-NOMA technique, while SU pairs communicate with each other through the STAR-RIS-assisted UAV. To maximize the sum rate of both PUs and SUs, the joint optimization problem of trajectory and transmission-reflection beamforming design of the STAR-RIS-assisted UAV, and power allocation is proposed. In addition, the formulated problem is the non-convex and NP-hard. Therefore, the alternative algorithm is obtained to approximate the original problem into three subproblems: i) the trajectory design of the STAR-RIS-assisted UAV by GP approximation, ii) the transmission-reflection beamforming design of the STAR-RIS-assisted UAV by SDR approach, and iii) the power allocation by D.C. method. The simulation results evaluated the objectives of the proposed system model across multiple categories. First, the impact of varying the important parameters on the performance, trajectory design of the STAR-RIS-assisted UAV, and interference management. Second, we explored how the dynamic environment affects trajectory design, throughput performance, and the energy splitting ratio of elements in the air-to-ground channel of the STAR-RIS-assisted UAV, considering both fixed and mobile PUs and SUs. Third, OMA is used as a benchmark compared to NOMA to examine the transmission technique's effectiveness on sum rate performance. Finally, to analyze the full coverage and high SE achievement in CR network, the different modes of intelligence surface mounted on the UAV. Also, the efficiency of various operating protocols such as ES, MS, and TS is illustrated.